\documentclass[pra,twocolumn,
superscriptaddress,
showpacs,
aps
]{revtex4-1}
\usepackage{graphicx}
\usepackage{amsmath}
\usepackage{amsfonts}
\usepackage{amssymb}
\usepackage{hyperref}
\usepackage{color}

\newcommand{\beq}{\begin{equation}}
	\newcommand{\eeq}{\end{equation}}

\usepackage{dsfont}
\usepackage{placeins} 
\usepackage{bm}           

\begin{document}
	\title{Collective emission of an atomic beam into an off-resonant cavity mode}
	\author{Simon~B.~J{\"a}ger}
	\affiliation{JILA, National Institute of Standards and Technology, and University of Colorado, Boulder, Colorado 80309-0440, USA}
	\author{Haonan~Liu}
	\affiliation{JILA, National Institute of Standards and Technology, and University of Colorado, Boulder, Colorado 80309-0440, USA}
	\author{John Cooper}
	\affiliation{JILA, National Institute of Standards and Technology, and University of Colorado, Boulder, Colorado 80309-0440, USA}
	\author{Murray~J.~Holland}
	\affiliation{JILA, National Institute of Standards and Technology, and University of Colorado, Boulder, Colorado 80309-0440, USA}
	
	\date{\today}
	
	\begin{abstract}
		We study the collective emission of a beam of atomic dipoles into an optical cavity. Our focus lies on the effect of a finite detuning between the atomic transition frequency and the cavity resonance frequency. By developing a theoretical description of the coupled atom-cavity dynamics we analyze the stationary atomic configurations including a superradiant phase where the atoms undergo continuous monochromatic collective emission. In addition, we derive an analytical formula for the cavity pulling coefficient which characterizes the displacement of the emission frequency towards the cavity frequency. We find that the pulling is small if the cavity linewidth is much larger than the collective linewidth of the atomic beam. This regime is desired for building stable lasers because the emission frequency is robust against cavity length fluctuations. Furthermore, we investigate the stability of the atomic phases and compare our theoretical predictions with numerical results. Remarkably, we also find polychromatic emission regimes, where the spectrum has several frequency components while the light output is still superradiant.
		
	\end{abstract}
	
	\maketitle
	
	\section{Introduction}
	Atomic ensembles in optical cavities provide a versatile platform to study collective effects that arise from strong light-matter interactions. These systems have been employed to study spatial pattern formation including self-organization~\cite{Domokos:2002,Baumann:2010,Ritsch:2013}, synchronization~\cite{Acebron:2005,Heinrich:2011,Xu:2014,Zhu:2015,Weiner:2017}, and also spin ordering or texturing~\cite{Mivehvar:2018,Kroeze:2018,Landini:2018}. They are intrinsically open quantum systems because photons can enter and leave through the cavity mirrors while external driving usually balances cavity losses and allows the stabilization of coherent out-of-equilibrium states. 
	
	The success of these systems also relies on the good controllability of cavity-mediated interactions in atomic systems. These can be tuned by adjusting the parameters of the driving lasers but also by varying the detunings between the atomic transitions and cavity mode frequencies. For instance, if an ensemble of metastable dipoles couple to a resonant cavity the dynamics will mostly be dominated by dissipation in form of spontaneous as well as superradiant or subradiant emission \cite{Dicke:1954,Gross:1982,Temnov:2005:1,Norcia:2016:2}. In contrast, for the case of large detuning, the dynamics remains coherent on long timescales, and these setups can be used for quantum simulations of collective physics~\cite{Muniz:2020} and even for spin squeezing~\cite{Leroux:2010,Swan:2018, Pedrozo:2020}.
	
	However, fluctuations in the cavity detuning are also a major source of noise. One of the main obstacles that limit the precision of the state-of-the-art cavity-assisted atomic clocks is the quantum noise caused by cavity detuning from mirror fluctuations. Recently, it has been found that the noise caused by such fluctuations can be minimized by having systems working in a  so-called `bad cavity' parameter regime~\cite{Meiser:2009,Bohnet:2012,Norcia:2016:1,Hotter:2019}. In this regime, the phase information of the output field is stored in the atomic ensemble rather than the cavity. Such systems, including active atomic clocks \cite{Chen:2009} and superradiant lasers \cite{Bohnet:2012,Norcia:2016:1,Laske:2019,Schaeffer:2020,Tang:2021,Debnath:2018}, are becoming candidates for future standards of quantum metrology. 
	
	Despite the fundamental interest in these kind of systems, only a few works investigate the effect of continuously introducing and removing atoms. Recently, the use of an atomic beam to study superradiant lasing and dynamical phases have been discussed~\cite{Temnov:2005:2,Liu:2020,Jaeger:2021:1,Jaeger:2021:2}. These atomic-beam cavity configurations represent interesting situations where neither photons nor individual atoms remain in the cavity on long timescales, but nevertheless cooperative effects can beat single-atom constraints. 
	
	In this paper we investigate the collective emission of an atomic beam into an off-resonant cavity. The finite detuning between the cavity and atomic transitions results in a collective Lamb shift~\cite{Gross:1982}. We investigate the special case where the atoms enter in their electronic excited state and discuss how the collectively emitted light depends on the detuning. We study cavity pulling effects in this setup, which describes the shift of the emission frequency in the direction of the cavity resonance, and investigate the dynamical superradiant phases that emerge. 
	
	This paper is structured as follows. In Sec.~\ref{sec:2} we introduce the theoretical framework to describe the coupled dynamics between the atomic beam and the cavity mode. We derive stationary solutions of this description in Sec.~\ref{sec:3} where we also derive an analytical expression for the cavity pulling coefficient. Section~\ref{sec:4} treats the stability of the stationary atomic configuration and studies the onset of superradiance and the destabilization of the superradiant phase. In Sec.~\ref{sec:5} we investigate a specific model and derive expressions for the stationary phases, and we compare our results to numerical simulations of this system. After that we conclude our results in Sec.~\ref{sec:6} while the Appendix provides further details to some calculations contained in Sec.~\ref{sec:4}.
	
	\section{\label{sec:2}Theoretical model}
	In this section we introduce the theoretical description of the dynamics of the atomic beam coupled to an off-resonant cavity.
	
	\subsection{\label{sec:master}Master equation formalism}
\begin{figure}[ht]
		\center
		\includegraphics[width=0.7\linewidth]{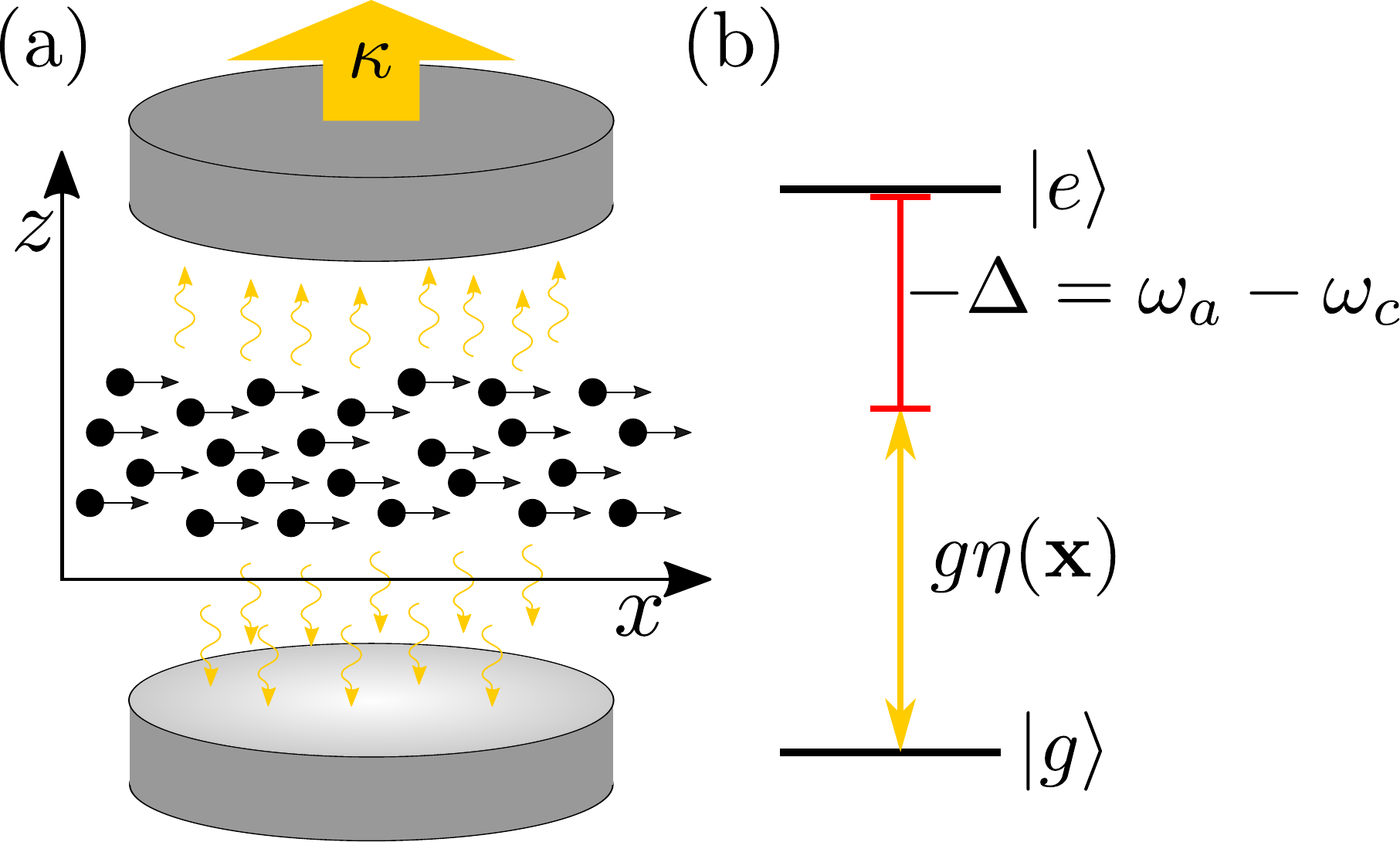}
		\caption{(a) Atoms enter the cavity in the excited state $|e\rangle$ and can emit photons into a single cavity mode. Photons leak out through the cavity output mirror with rate $\kappa$. (b) Each atom is represented as an optical dipole of transition frequency $\omega_a$ coupled to the cavity mode of frequency $\omega_c$. The coupling $g\eta({\bf x})$ depends on the position ${\bf x}$ of the atom, where $g$ is the vacuum Rabi frequency or Jaynes-Cummings coupling coefficient and $\eta({\bf x})$ is the mode function. The cavity-atom detuning frequency is given by $\Delta = \omega_c - \omega_a$. \label{Fig:1}}
	\end{figure}

	We consider a beam of two-level atomic dipoles in their excited state $|e\rangle$ with transition frequency $\omega_a$ and mass~$m$ traversing an optical cavity. These atoms can emit photons into a single cavity mode of frequency $\omega_c$ (see Fig.~\ref{Fig:1}). 
	The density matrix $\hat{\rho}$ describing the atomic and cavity degrees of freedom is governed by a Born-Markov master equation
	\begin{align}
		\label{Mastereq}
		\frac{d \hat{\rho}}{d t}=\frac{1}{i\hbar}\left[\hat{H},\hat{\rho}\right]-\frac{\kappa}{2}(\hat{a}^{\dag}\hat{a}\hat{\rho}+\hat{\rho}\hat{a}^{\dag}\hat{a}-2\hat{a}\hat{\rho}\hat{a}^{\dag}).
	\end{align}
	Here, the Hamiltonian 
	\begin{align}
		\label{H}
		\hat{H}=\hbar \Delta\hat{a}^{\dag}\hat{a}+\sum_{j}\left[\frac{\hat{\bf p}_j^2}{2m}+\frac{\hbar g}{2}\eta(\hat{\bf x}_j)\left(\hat{a}^{\dag}\hat{\sigma}_j^-+\hat{\sigma}_j^{+}\hat{a}\right)\right]
	\end{align}
	describes the coherent dynamics of the coupled atom-cavity system in a frame rotating with $\omega_a$. The first term determines the energy of cavity photons where $\Delta=\omega_c-\omega_a$ is the detuning between the cavity and the atomic frequency. Operators $\hat{a}$ and $\hat{a}^{\dag}$ are the photonic annihilation and creation operators that fulfill the commutation relation $[\hat{a},\hat{a}^{\dag}]=1$. The second term in Eq.~\eqref{H} is the kinetic energy of atom $j$ where $j$ runs over all atoms in the atomic beam. The last term in Eq.~\eqref{H} describes the Jaynes-Cummings coupling between an atom and the cavity, where $g$ is the vacuum Rabi frequency at a field maximum and $\eta({\bf \hat{x}})$ is the cavity mode function evaluated at position $\hat{\bf x}$. 
	
	The atomic position operator
	${\hat{\bf x}_j=(\hat{x}_j,\hat{y}_j,\hat{z}_j)^T}$ is conjugate to the momentum operator ${\hat{\bf p}_j=(\hat{p}_{x,j},\hat{p}_{y,j},\hat{p}_{z,j})^T}$ with the usual canonical commutation relations $[\hat{\mu}_j,\hat{p}_{\nu,k}]=i\hbar\delta_{jk}\delta_{\mu\nu}$, $\mu,\nu\in\{x,y,z\}$. The operators $\hat{\sigma}_j^+=|e\rangle_j\langle g|_j$ and $\hat{\sigma}_j^-=|g\rangle_j\langle e|_j$ are the atomic raising and lowering operators, where $|e\rangle_j$, $|g\rangle_j$ denote electronic excited and ground states.
	
	Dissipation in this system is described by the Lindblad term in the master equation Eq.~\eqref{Mastereq}. This describes the leakage of cavity photons into the free-space electromagnetic field with rate $\kappa$, typically referred to as the cavity linewidth.
	
	\subsection{Heisenberg-Langevin equations}
	The master equation formalism introduced in Sec.~\ref{sec:master} is equivalent to the Heisenberg-Langevin equations that are given by
	\begin{align}
		\label{a}
		\frac{d\hat{a}}{dt}=&-\left(i\Delta +\frac{\kappa}{2}\right)\hat{a}-i\frac{g}{2}\hat{J}^{-}+\hat{\mathcal{F}}^{-},\\
		\label{sigma-}
		\frac{d\hat{\sigma}_j^{-}}{dt}=&\frac{ig}{2}\eta(\hat{\bf x}_j)\hat{\sigma}_j^{z}\hat{a},\\
		\label{sigmaz}
		\frac{d\hat{\sigma}_j^{z}}{dt}=&ig\eta(\hat{\bf x}_j)\left(\hat{a}^{\dag}\hat{\sigma}^-_j-\hat{\sigma}^{+}_j\hat{a}\right),\\
		\label{hatx}
		\frac{d\hat{\bf x}_j}{dt}=&\frac{\hat{\bf p}_j}{m},\\
		\label{hatp}
		\frac{d\hat{\bf p}_j}{dt}=&-\frac{g\hbar}{2}(\hat{a}^{\dag}\hat{\sigma}_j^{-}+\hat{\sigma}_j^{+}\hat{a})\left.\nabla_{\bf x}\eta({\bf x})\right|_{{\bf x}=\hat{\bf x}_j}.
	\end{align}
	Here we have represented the gradient as $\nabla_{\bf x}\equiv(\partial/\partial x,\partial/\partial y,\partial/\partial z )^T$, and included the cavity shot noise~$\hat{\mathcal{F}}^{-}$ that fulfills expectation values $\langle \hat{\mathcal{F}}^{-}(t)\rangle=0$, $\langle \hat{\mathcal{F}}^{-}(t')\hat{\mathcal{F}}^{-}(t)\rangle=0=\langle \hat{\mathcal{F}}^{+}(t')\hat{\mathcal{F}}^{-}(t)\rangle$, and $\langle \hat{\mathcal{F}}^{-}(t')\hat{\mathcal{F}}^{+}(t)\rangle=\kappa\delta(t-t')$, with $\hat{\mathcal{F}}^{+}=(\hat{\mathcal{F}}^{-})^{\dag}$. The operators $\hat{\sigma}_j^{z} = \hat{\sigma}_j^{+}\hat{\sigma}_j^{-} - \hat{\sigma}_j^{-}\hat{\sigma}_j^{+}$ describe the population inversion. The operator $\hat{J}^-$ is the collective dipole and is defined as
	\begin{align}
		\hat{J}^-=\sum_j\eta(\hat{\bf x}_j)\hat{\sigma}_j^{-}.
	\end{align}
	
	We are interested in the situation where dipoles in the atomic beam transverse the cavity mode with a large velocity. Assuming a mean velocity $v_x$ perpendicular to the cavity axis [see Fig.\ref{Fig:1}(a)], we can estimate the transit time as $\tau=2w/v_x$ where $w$ is the beam waist of the cavity mode. Throughout this paper we will neglect optomechanical forces that are described by Eq.~\eqref{hatp} and consider only ballistic motion. This is valid in the parameter regime where the atomic momentum distribution has a width $\Delta p_\mu=\sqrt{\langle\hat{p}_\mu^2\rangle-\langle\hat{p}_\mu\rangle^2}$ that exceeds the mean force $F_a$ times the transit time $\tau$ in every spatial direction $\mu\in\{x,y,z\}$. In this regime we may assume that the momentum of each atom is constant. 
	
	\subsection{Semiclassical description}
	We will now make a semiclassical approximation where we substitute the operators by $c$-number variables and add noise terms that give the correct second moments. Similar approaches have been used in Refs.~\cite{Liu:2020,Jaeger:2021:1,Jaeger:2021:2,Schachenmayer:2015}. Specifically, we replace the position operators $\hat{\bf x}_j$ by the classical variables ${\bf x}_j$. We derive the time evolution of the Hermitian cavity operators $\hat{a}^x=\hat{a}+\hat{a}^{\dag}$, $\hat{a}^y=i(\hat{a}-\hat{a}^{\dag})$ and atomic dipole operators $\hat{\sigma}^x_j=\hat{\sigma}^{-}_j+\hat{\sigma}^{+}_j$, $\hat{\sigma}^y_j=i(\hat{\sigma}^{-}_j-\hat{\sigma}^{+}_j)$, $\hat{\sigma}^z_j$, and then substitute them by their classical counterparts; $\alpha^x$, $\alpha^y$ for the cavity and $s^x_j$, $s^y_j$, and $s^z_j$ for the dipoles. The $c$-number noise terms are chosen such that the second moments of two classical variables $A$, $B$ relate in the form $\langle AB\rangle=\langle \hat{A}\hat{B}+\hat{B}\hat{A}\rangle/2$ to the second moment of their corresponding operators $\hat{A}$ and $\hat{B}$; i.e., we choose symmetric ordering of the operators. The resulting $c$-number stochastic differential equations read
	\begin{align}
		\label{alphax}
		\frac{d\alpha^x}{dt}=&-\frac{\kappa}{2}\alpha^x-\Delta\alpha^y-\frac{g}{2}J^y+\mathcal{F}^x,\\
		\label{alphay}
		\frac{d\alpha^y}{dt}=&\Delta\alpha^x-\frac{\kappa}{2}\alpha^y+\frac{g}{2}J^x+\mathcal{F}^y,\\
		\label{sx}
		\frac{ds^x_j}{dt}=&\frac{g}{2}\eta({\bf x}_j)s^z_j\alpha^y,\\
		\label{sy}
		\frac{ds^y_j}{dt}=&-\frac{g}{2}\eta({\bf x}_j)s^z_j\alpha^x,\\
		\label{sz}
		\frac{ds^z_j}{dt}=&\frac{g}{2}\eta({\bf x}_j)(\alpha^xs^y_j-\alpha^ys^x_j).\\
		\label{x}
		\frac{d{\bf x}_j}{dt}=&\frac{{\bf p}_j}{m},
	\end{align}
	Here, $\mathcal{F}^x$ and $\mathcal{F}^y$ are independent noise terms defined by $\langle\mathcal{F}^x\rangle=\langle\mathcal{F}^y\rangle=\langle\mathcal{F}^x(t)\mathcal{F}^y(t')\rangle=0$ and $\langle\mathcal{F}^x(t)\mathcal{F}^x(t')\rangle=\langle\mathcal{F}^y(t)\mathcal{F}^y(t')\rangle=\kappa\delta(t-t')$. In Eqs.~\eqref{alphax}--\eqref{alphay}, $J_x$ and $J_y$ are the classical $x$ and $y$ components of the collective dipole given by $J^x=\sum_j\eta({\bf x}_j)s_j^{x}$ and $J^y=\sum_j\eta({\bf x}_j)s_j^{y}$. Eq.~\eqref{x} describes the ballistic trajectory.
	
	Noise is not only introduced by the cavity degrees of freedom, but also by the boundary conditions. We will investigate the dynamics of atoms that enter the cavity in the excited state $|e\rangle$. Therefore if an atom indexed by $j$ enters the cavity, we initialize $s_j^z=1$ and choose the $x$ and $y$ components of the dipoles randomly and independently from $s_j^x=\pm1$ and $s_j^y=\pm1$. This accounts for the correct second moments of all dipole components (see Refs.~\cite{Liu:2020,Jaeger:2021:1,Jaeger:2021:2}).
	With these boundary constraints, Eqs.~\eqref{alphax}--\eqref{x} can be directly implemented in numerical simulations. In the next section, we will introduce a density method to analytically solve these equations. 
	
	\subsection{Density description}
	We will now use Eqs.~\eqref{sx}--\eqref{x} to derive a collective description of the atomic beam. For this we define the densities
	\begin{align}
		f({\bf x},{\bf p},t)=&\sum_{j}\delta({\bf x}-{\bf x}_j)\delta({\bf p}-{\bf p}_j),\\
		s^{\mu}({\bf x},{\bf p},t)=&\sum_{j}s_j^{\mu}\delta({\bf x}-{\bf x}_j)\delta({\bf p}-{\bf p}_j),
	\end{align}
	where the $s_j^{\mu}$ are spin components with $\mu\in\{x,y,z\}$. Using these definitions together with Eqs.~\eqref{sx}--\eqref{x} we obtain
	\begin{align}
		\label{f}
		\frac{\partial f}{\partial t}+\frac{\bf p}{m}\cdot \nabla_{\bf x}f=&0,\\
		\label{sxdensity}
		\frac{\partial s^x}{\partial t}+\frac{\bf p}{m}\cdot \nabla_{\bf x}s^x=&\frac{g}{2}\eta({\bf x})s^z\alpha^y,\\
		\label{sydensity}
		\frac{\partial s^y}{\partial t}+\frac{\bf p}{m}\cdot \nabla_{\bf x}s^y=&-\frac{g}{2}\eta({\bf x})s^z\alpha^x,\\
		\label{szdensity}
		\frac{\partial s^z}{\partial t}+\frac{\bf p}{m}\cdot \nabla_{\bf x}s^z=&\frac{g}{2}\eta({\bf x})\left(\alpha^{x}s^{y}-\alpha^{y}s^{x}\right).
	\end{align}
	The collective dipole in Eqs.~\eqref{alphax}--\eqref{alphay} can be also expressed as an integral over dipole densities
	\begin{align}
		\label{colldipole}
		J^\mu=\int d{\bf x}\int d{\bf p}\,\eta({\bf x})s^\mu({\bf x},{\bf p},t),
	\end{align}
	with $\mu\in\{x,y\}$. 
	
	Equations~\eqref{f}--\eqref{szdensity} are closed with the time evolution of the field variables in Eq.~\eqref{alphax}--\eqref{alphay}.	It remains to include the atomic noise terms in this density formalism. To do this, we formulate the initial conditions for the atoms entering the cavity as boundary conditions for the partial differential equations~\eqref{f}--\eqref{szdensity}. 
	Assuming that the atoms enter the cavity in the plane $x=-x_0$ [see Fig.~\ref{Fig:1}(a)], we can ascribe as initial conditions;
	\begin{align}
		f(-x_0,y,z,{\bf p},t)=&f_0(y,z,{\bf p},t),\\
		s^x(-x_0,y,z,{\bf p},t)=&W^x(y,z,{\bf p},t),\\
		s^y(-x_0,y,z,{\bf p},t)=&W^y(y,z,{\bf p},t),\\
		s^z(-x_0,y,z,{\bf p},t)=&f_0(y,z,{\bf p},t).
	\end{align}
	The boundary condition for the density is given by 
	\begin{align}
		f_0(y,z,{\bf p},t)=\sum_j\delta({\bf x}_0-{\bf x}_j)\delta({\bf p}-{\bf p}_j),
	\end{align}
	where ${{\bf x}_0=(-x_0,y,z)^T}$ is the position where the atoms enter. We can therefore express the initial condition for the dipoles as
	\begin{align}
		W^\mu(y,z,{\bf p},t)=\sum_js_j^\mu\delta({\bf x}_0-{\bf x}_j)\delta({\bf p}-{\bf p}_j),
	\end{align}
	with $\mu\in\{x,y\}$, and for the second moment as
	\begin{align}
		\label{noise}
		\langle W^{\mu}(W^{\nu})'\rangle = &\frac{m}{p_x}\delta_{\mu\nu}\delta(t-t')\delta(y-y')\delta(z-z') \notag\\
		&\times \delta({\bf p}-{\bf p}')f_0(y,z,{\bf p},t),
	\end{align}
	where we have used the notation $W^{\mu}=W^{\mu}(y,z,{\bf p},t)$ and $(W^{\nu})'=W^{\nu}(y',z',{\bf p}',t')$.
	
	Throughout this paper we will assume that the atomic density after one transit time $\tau$ is spatially homogeneous in the cavity. This results in the property
	\begin{align}
		\langle f({\bf x},{\bf p},t\gg\tau)\rangle=\rho({\bf p}),
	\end{align}
	where $\rho({\bf p})$ is a continuous spatially homogeneous density of the atoms that is time independent. This, however, does not imply that the dipole densities $s^\mu$, $\mu\in\{x,y,z\}$, are spatially independent, as we will expand on in the next section.
	
	\section{\label{sec:3}Stationary states of the system}
	We will now investigate the asymptotic stationary solution reached after a sufficiently long time, $t \gg \tau$, of the coupled equations for the field, Eqs.~\eqref{alphax}--\eqref{alphay}, and dipole densities, Eqs.~\eqref{sxdensity}--\eqref{szdensity}. To obtain these results we will use the previously mentioned assumption of a spatially homogeneous atomic density. In addition, we will discard all noise terms, which implies a mean-field approximation.
	
	\subsection{The non-superradiant solution}
	We begin with the simplest solution that describes the situation when the atoms cross the cavity without generating a coherent light field. This is a trivial stationary state of the system given by
	\label{nonsupstate}
	\begin{align}
		\alpha^x_0=&0,\nonumber\\
		\alpha^y_0=&0,\nonumber\\
		s^x_0=&0,\nonumber\\
		s^y_0=&0,\nonumber\\
		s^z_0=&\rho({\bf p}).
	\end{align}
	In this case, the atoms simply remain in the excited state $|e\rangle$ while traveling through the cavity region. 
	
	\subsection{The superradiant solution}
	We now derive the more interesting superradiant solution. In order to reduce the equations, we rotate to a complex field ${\alpha=(\alpha^x-i\alpha^y)/2}$ and complex dipole density ${s=(s^x-is^y)/2}$. Using these definitions in Eqs.~\eqref{alphax}--\eqref{alphay} and Eqs.~\eqref{sxdensity}--\eqref{sydensity}, we derive the following mean-field equations
	\begin{align}
		\label{malpha}
		\frac{d\alpha}{d t}=&-\left(i\Delta+\frac{\kappa}{2}\right)\alpha-i\frac{g}{2}\int d{\bf x}\int d{\bf p}\,\eta({\bf x})s,\\
		\label{mfs}
		\frac{\partial s}{\partial t}=&-\frac{\bf p}{m}\cdot\nabla_{\bf x}s+\frac{ig}{2}\eta ({\bf x})s^z\alpha,\\
		\label{mfz}
		\frac{\partial s^{z}}{\partial t}=&-\frac{\bf p}{m}\cdot\nabla_{\bf x}s^z+ig\eta({\bf x})\left(\alpha^*s-s^*\alpha\right),
	\end{align}
	where we have used the collective dipole defined in Eq.~\eqref{colldipole}.
	
\begin{figure}[h!]
		\center
		\includegraphics[width=0.5\linewidth]{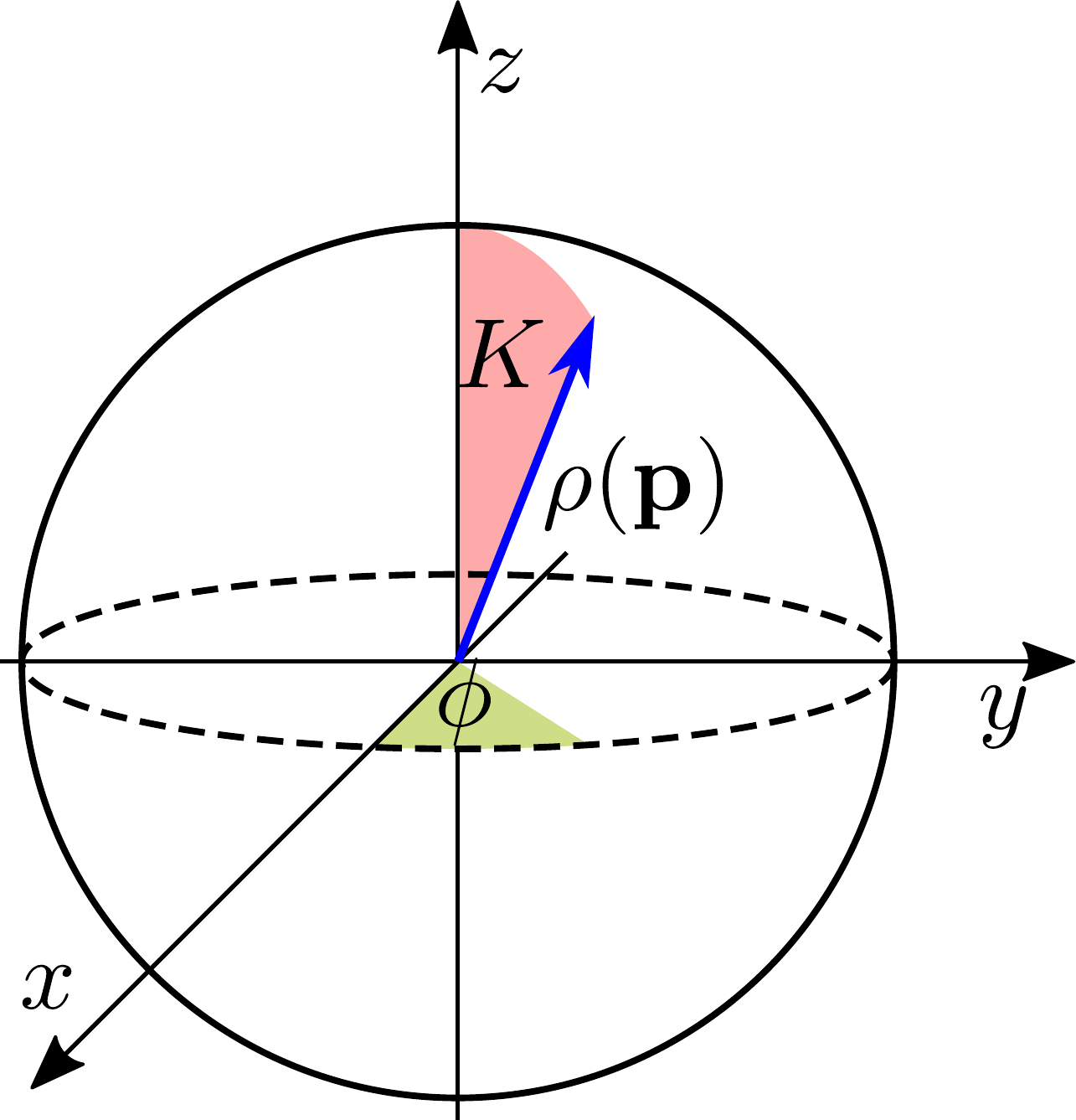}
		\caption{Sketch of the Bloch sphere where the dipole density can be mapped on a point of the sphere (here visible as the blue arrow) with radius $\rho({\bf p})$ depending solely on the momentum~${\bf p}$. The angles $K$ and $\phi$ depend on position ${\bf x}$, momentum~${\bf p}$, and time $t$.\label{Fig:2}}
	\end{figure}

	Equations~\eqref{mfs}--\eqref{mfz} imply a conserved length of the dipole density
	\begin{align}
		\left(\frac{\partial}{\partial t}+\frac{\bf p}{m}\cdot\nabla_{\bf x}\right)\left[4|s|^2+(s^z)^2\right]=0,
	\end{align}
	which can be seen by realizing $4|s|^2+(s^z)^2=(s^x)^2+(s^y)^2+(s^z)^2$. Therefore it is useful to represent the stationary dipole components in spherical coordinates 
	\begin{align}
		s_0=&\frac{\rho({\bf p})}{2}e^{-i\phi}\sin(K),\nonumber\\
		s^z_0=&\rho({\bf p})\cos(K),
		\label{Superradiantsolution}
	\end{align}
	where the dipole length is determined by the boundary conditions of the atomic beam density $\rho({\bf p})$ as in Eq.~\eqref{f}, and $\phi$, $K$ are spherical angles dependent on position, momentum, and time. In that case, for every fixed value of ${\bf x}$, ${\bf p}$, $t$, we can assign a Bloch vector to the density of the atomic dipoles (see Fig.~\ref{Fig:2}). 
	The boundary condition for $K$ is determined by the fact that the atoms enter in the excited state and thus $K({\bf x}_0,{\bf p},t)=0$.
	
	To find the superradiant solution, we assume that the atomic beam undergoes collective emission with a single frequency $\omega$. In that case we can express the phase $\phi$ as
	\begin{align}
		\phi({\bf x},{\bf p},t)=\omega t+\psi({\bf x},{\bf p}),
	\end{align}
	where the first term on the right hand side describes the monochromatic oscillation of the density with frequency~$\omega$, and the second term $\psi$ is a time-independent phase in phase space. The angle $K({\bf x},{\bf p})$ is not explicitly time dependent in this case.
	
	This assumption allows us to solve the cavity field analytically from Eq.~\eqref{malpha} and obtain
	\begin{align}
		\alpha_0 \approx -i\frac{\Gamma_c}{g}\cos(\chi)e^{-i\chi}J_0,\label{eliminatedalpha}
	\end{align}
	where we have defined
	\begin{align}
		\Gamma_c=&\frac{g^2}{\kappa},\\
		\label{tanchi}
		\tan(\chi)=&\frac{\Delta-\omega}{\kappa/2},
	\end{align}
	and
	\begin{align}
		\label{J0}
		J_0=\int d{\bf x}\int d{\bf p}\eta({\bf x})s_0.
	\end{align}
	We mention that $\alpha_0$, $s_0$, and $J_0$ are all proportional to $\exp(-i\omega t)$, which constitutes their only time dependence. Our result for the field goes beyond the typical adiabatic elimination of the cavity fields since it includes retardation effects that are apparent in $\chi$ and that explicitly depend on the frequency $\omega$. Using Eqs.~\eqref{Superradiantsolution} and Eq.~\eqref{eliminatedalpha} in Eq.~\eqref{mfs}, we can derive the following equations for the angles
	\begin{align}
		\label{1}
		\frac{\bf p}{m}\cdot\nabla_{\bf x}\psi=&-\omega-\frac{\Gamma({\bf x})}{2}\cot(K)C(\psi),\\
		\label{2}
		\frac{\bf p}{m}\cdot\nabla_{\bf x}K=&\frac{\Gamma({\bf x})}{2}\int d{\bf x}'S(\psi),
	\end{align}
	with
	\begin{align*}
		C(\psi)=&\int d{\bf x}'\int d{\bf p}'\,\eta'\rho'\sin\left(\psi-\psi'-\chi\right)\sin(K'),\\
		S(\psi)=&\int d{\bf p}'\,\eta'\rho'\cos\left(\psi-\psi'-\chi\right)\sin(K'),
	\end{align*}
	and where we have used
	\begin{align}
		\Gamma({\bf x})=\Gamma_c\eta({\bf x})\cos(\chi). \label{Gamma(x)}
	\end{align} 
	As a simplification, we have employed the notation $\mathcal{A}'=\mathcal{A}({\bf x}',{\bf p}')$ where $\mathcal{A}$ can be $\eta$, $\rho$, $\psi$, and $K$. Equations~\eqref{1}--\eqref{2} have a $U(1)$ symmetry since they are invariant under a rotation $\psi\mapsto \psi+\varphi$, where $\varphi$ is an arbitrary phase that is independent on position, momentum, and time. We will now explicitly break this $U(1)$ symmetry by choosing the phase offset such that
	\begin{align}
		\label{Jpar0}
		J_0^{\parallel}=&\int d{\bf x}'\int d{\bf p}'\,\eta'\rho'\cos\left(\psi'\right)\sin(K'),\\
		\label{zero}
		0=&\int d{\bf x}'\int d{\bf p}'\,\eta'\rho'\sin\left(\psi'\right)\sin(K').
	\end{align}
	Notice that $J_0^{\parallel}$ is not time dependent; the value of $J_0^{\parallel}$ is the stationary length of the collective dipole and has the relation $J_0^{\parallel}=2|J_0|$.
	
	With this choice of $J_0^{\parallel}$, we can simplify Eq.~\eqref{1} and Eq.~\eqref{2} to
	\begin{align}
		\label{psi}
		\frac{\bf p}{m}\cdot\nabla_{\bf x}\psi=&-\omega-\frac{\Gamma({\bf x})J_0^{\parallel}}{2}\cot(K)\sin\left(\psi-\chi\right),\\
		\label{K}
		\frac{\bf p}{m}\cdot\nabla_{\bf x}K=&\frac{\Gamma({\bf x})J_0^{\parallel}}{2}\cos\left(\psi-\chi\right).
	\end{align}
	Since all atoms enter the cavity in the excited state we have the boundary condition $K({\bf x}_0,{\bf p})=0$. If we now impose that the gradient of the angle $\nabla_{\bf x}\psi$ cannot diverge at ${\bf x}={\bf x}_0$, we obtain the boundary condition for the angle $\psi({\bf x}_0,{\bf p})=\chi$. 
	
	Although we will give a simple example in Sec.~\ref{sec:5} where we can explicitly solve Eqs.~\eqref{psi}--\eqref{K}, we are not aware of a general solution. However, in the limit where $\chi\ll1$, we can apply perturbation theory as we will show now.
	
	\subsection{Perturbative solution for $\chi\ll1$: Cavity pulling}
	We consider now the case where $\chi\ll1$ and also $\psi\ll1$. The latter is a consequence of the boundary condition $\psi({\bf x}_0,{\bf p})=\chi$ together with the approximation $\sin\left(\psi-\chi\right)\approx\psi-\chi$ that implies that $\psi$ according to Eq.~\eqref{psi} is only slowly varying. In this parameter regime we can approximate $\chi$ by
	\begin{align}
		\label{approxchi}
		\chi\approx\frac{\Delta-\omega}{\kappa/2}
	\end{align}
	from Eq.~\eqref{tanchi} and simplify Eq.~\eqref{psi} and Eq.~\eqref{K} to get
	\begin{align}
		\label{psi2}
		\frac{\bf p}{m}\cdot\nabla_{\bf x}\psi=&-\omega-\frac{\Gamma_c J_0^{\parallel}}{2}\eta\cot(K)\left(\psi-\chi\right),\\
		\label{K2}
		\frac{\bf p}{m}\cdot\nabla_{\bf x}K=&\frac{\Gamma_c J_0^{\parallel}}{2}\eta.
	\end{align}
	The second equation is now completely decoupled and independent of $\omega$. Using the substitution
	\begin{align}
		\label{psi3}
		\psi=\frac{\Psi}{\sin(K)}+\chi,
	\end{align}
	we can derive
	\begin{align}
		\frac{\bf p}{m}\cdot\nabla_{\bf x}\Psi=-\omega\sin(K)
	\end{align}
	with the boundary condition $\Psi({\bf x}_0,{\bf p})=0$. This can be integrated to obtain
	\begin{align}
		\label{Psiint}
		\Psi({\bf x},{\bf p})=-\omega\int_{0}^{\infty}dt\,\sin\left[K\left({\bf x}-\frac{\bf p}{m}t,{\bf p}\right)\right],
	\end{align}
	where we have extended the upper limit of the integral to infinity assuming that $K(x,y,z,{\bf p})=0$ for $x<-x_0$.
	
	Using Eq.~\eqref{psi3} and Eq.~\eqref{Psiint} in Eq.~\eqref{Jpar0} and Eq.~\eqref{zero}, we obtain
	\begin{align}
		\label{Jpar0Psi}
		0=&\int d{\bf x}'\int d{\bf p}'\,\eta'\rho'\Psi'+\chi J_0^{\parallel}.
	\end{align}
	Combining Eq.~\eqref{approxchi}, Eq.~\eqref{Psiint}, and Eq.~\eqref{Jpar0Psi}, we can now solve for the frequency 
	\begin{align}
		\omega=\frac{\Delta}{\frac{\kappa C_{\perp}}{2}+1}\label{frequency}
	\end{align} 
	where we have defined
	\begin{align}
		\label{Cperp}
		C_{\perp}=&\frac{\int_{0}^{\infty} dt\int d{\bf x}\int d{\bf p}\eta\left({\bf x}+\frac{\bf p}{m}t\right)\rho\sin(K)}{J^\parallel_0}
	\end{align} 
	as a timescale.
	
	The result given in Eq.~\eqref{frequency} can be rewritten to calculate the cavity pulling coefficient 
	\begin{align}
		\label{Pulling}
		\mathcal{P}=\frac{\omega}{\Delta}=\frac{1}{\frac{\kappa C_{\perp}}{2}+1}
	\end{align}
	that describes the emission frequency of the atomic beam relative to the detuning between the cavity resonance and the atomic resonance. While the exact form of $C_{\perp}$ depends on the actual model, there is still a very general physical observation that we can make. If the timescale $C_{\perp}$ is small enough such that $\kappa C_{\perp}\ll1$, we get a pulling coefficient $\mathcal{P}\lesssim1$. In this case, light will essentially be emitted with the cavity frequency and not with the atomic frequency for $\Delta \ll \kappa$. On the other hand, if $\kappa C_{\perp}\gg1$, we have a cavity pulling coefficient $\mathcal{P}\ll1$ and therefore the emitted light is almost resonant with the atomic transition frequency. This has been shown to be the case for superradiant lasers~\cite{Bohnet:2012, Liu:2020} that work in the regime where $\kappa$ is much larger than any atomic linewidth, in particular $\kappa\gg N\Gamma_c$. For situations where a stable emission frequency is desired that is independent of cavity length noise, we would like $\mathcal{P}$ to be as small as possible. For the remainder of this article we will now focus exactly on this regime and first determine the stability of the atomic beam configuration. 
	
	\section{\label{sec:4}Stability in the bad cavity regime} 
	In the limit where $\kappa$ determines the shortest timescale, we can eliminate $\alpha$ from Eqs.~\eqref{sxdensity}--\eqref{szdensity} according to Eq.~\eqref{eliminatedalpha} and also neglect the the explicit $\omega$ dependence of $\chi$, i.e.,
	\begin{equation}
		\tan(\chi)=\frac{\Delta}{\kappa/2}.
	\end{equation} 
	We then obtain the following stochastic differential equations for the dipole densities
	\begin{align}
		\label{sxdensityel}
		\frac{\partial s^x}{\partial t}+\frac{\bf p}{m}\cdot \nabla_{\bf x}s^x=&\frac{\Gamma({\bf x})}{2}\left[\cos(\chi)J^{x}-\sin(\chi)J^{y}\right]s^z+\mathcal{S}^x,\\
		\label{sydensityel}
		\frac{\partial s^y}{\partial t}+\frac{\bf p}{m}\cdot \nabla_{\bf x}s^y=&\frac{\Gamma({\bf x})}{2}\left[\sin(\chi)J^{x}+\cos(\chi)J^{y}\right]s^z+\mathcal{S}^y,\\
		\frac{\partial s^z}{\partial t}+\frac{\bf p}{m}\cdot \nabla_{\bf x}s^z=&-\frac{\Gamma({\bf x})}{2}\cos(\chi)\left(s^{x}J^{x}+s^{y}J^{y}\right) \notag\\
		\label{szdensityel}
		&-\frac{\Gamma({\bf x})}{2}\sin(\chi)\left(s^{y}J^{x}-s^{x}J^{y}\right) +\mathcal{S}^z,
	\end{align}
	where we have used the definition given in Eq.~\eqref{Gamma(x)}.
	
	Equations~\eqref{sxdensityel}--\eqref{szdensityel} also include stochastic noise terms $\mathcal{S}^x=\eta({\bf x})\mathcal{N}^xs^z$, $\mathcal{S}^y=\eta({\bf x})\mathcal{N}^ys^z$, and $\mathcal{S}^z=-\eta({\bf x})\left(\mathcal{N}^xs^x+\mathcal{N}^ys^y\right)$, where the noise terms $\mathcal{N}^x$ and $\mathcal{N}^y$ can be assumed to be delta-correlated on the typical evolution timescale of the atomic degrees of freedom. This implies ${\langle\mathcal{N}^x(t)\mathcal{N}^y(t')\rangle=0}$ and  ${\langle\mathcal{N}^x(t)\mathcal{N}^x(t')\rangle=\langle\mathcal{N}^y(t)\mathcal{N}^y(t')\rangle}=\Gamma_c\cos^2(\chi)\delta(t-t')$.
	
	These noise terms are important for the dynamics since they introduce small fluctuations into the dipole components that can destabilize the state. In order to predict this destabilization, we investigate the stability of the stationary phases that we have introduced in Sec.~\ref{sec:3}.
	
	\subsection{Stability of the non-superradiant configuration}
	For the non-superradiant configuration, we study small fluctuations $\delta s^x$ and $\delta s^y$ around the solution given in Eq.~\eqref{nonsupstate}. For this kind of analysis we can drop the noise terms. We find the linearized equations
	\begin{align}
		\label{deltasx}
		\frac{\partial \delta s^x}{\partial t}+\frac{\bf p}{m}\cdot \nabla_{\bf x}\delta s^x=&\frac{\Gamma({\bf x})}{2}\left[\cos(\chi)\delta J^{x}-\sin(\chi)\delta J^{y}\right]\rho,\\
		\label{deltasy}
		\frac{\partial \delta s^y}{\partial t}+\frac{\bf p}{m}\cdot \nabla_{\bf x}\delta s^y=&\frac{\Gamma({\bf x})}{2}\left[\sin(\chi)\delta J^{x}+\cos(\chi)\delta J^{y}\right]\rho.
	\end{align}
	Since we neglect terms that are second-order in the fluctuations, these equations become decoupled from fluctuations $\delta s^z$ around $s^z=\rho$ . We have also introduced $\delta J^\mu=\int d{\bf x}\int d{\bf p}\eta\delta s^\mu$  with $\mu=x,y$. 
	
	Equations~\eqref{deltasx}--\eqref{deltasy} can be reduced to uncoupled equations for $\delta s=(\delta s^x-i\delta s^y)/2$ and its complex conjugate. Without loss of generality, we focus on the solution of $\delta s$ and derive
	\begin{align}
		\frac{\partial \delta s}{\partial t}+\frac{\bf p}{m}\cdot \nabla_{\bf x}\delta s=&\frac{\Gamma({\bf x})}{2}e^{-i\chi}\delta J\rho,\label{deltas}
	\end{align}
	where $\delta J=(\delta J^x-i\delta J^y)/2$. Applying the Laplace transformation
	\begin{align}
		\label{laplace}
		L[g](\nu)=\int_{0}^{\infty}\,dt\,e^{-\nu t}g(t)
	\end{align}
	to Eq.~\eqref{deltas}, we obtain
	\begin{align}
		\label{nuL0}
		\left[\nu-\mathcal{L}_0\right]L[\delta s]=\delta s({\bf x},{\bf p},0)+\frac{\Gamma({\bf x})}{2}e^{-i\chi}\rho({\bf p})L[\delta J],
	\end{align}
	where we have defined the operator
	\begin{align}
		\label{L0}
		\mathcal{L}_0g({\bf x})=-\frac{\bf p}{m} \cdot \nabla_{\bf x}g({\bf x}).
	\end{align}
	Multiplying Eq.~\eqref{nuL0} by the inverse of operator $\left[\nu-\mathcal{L}_0\right]$ and $\eta({\bf x})$, and then integrating over space and momentum, we obtain 
	\begin{align}
		L[\delta J]=\frac{\int d{\bf x}\int d{\bf p}\eta({\bf x})\left[\nu-\mathcal{L}_0\right]^{-1}\delta s^x({\bf x},{\bf p},0)}{D(\nu)},
	\end{align}
	where the denominator is given by the dispersion relation
	\begin{align}
		\label{Dispersionrelationnonsupp}
		D(\nu) =&1-\int_{0}^{\infty} dt e^{-\nu t-i\chi}\int d{\bf x}\int d{\bf p}\eta\left({\bf x}+\frac{\bf p}{m}t\right)\frac{\Gamma({\bf x})}{2}\rho.
	\end{align}
	
	The asymptotic time evolution of $\delta J$ is determined by the zeros of the dispersion relation $D(\nu)$. In fact the zero,~$\nu_0$, that has the largest real component is the principal one that controls the dynamics. As long as we satisfy $\mathrm{Re}(\nu_0)<0$, the non-superradiant configuration is stable. The imaginary part $\mathrm{Im}(\nu_0)$ then determines the frequency of the light emission.
	
	In the case where $\mathrm{Re}(\nu_0)>0$, a qualitatively distinct solution is anticipated in which we expect an exponential build-up of fluctuations that results in superradiant emission, implying the formation of a macroscopic collective dipole. In the remainder of this section, we will determine the stability of this stationary superradiant phase.
	
	\subsection{Stability of the superradiant configuration}
	We analyze the dynamics of small fluctuations around the configuration that is determined by Eq.~\eqref{Superradiantsolution}, Eq.~\eqref{psi2}, and Eq.~\eqref{K2}. To do so it is convenient to move into a frame rotating with frequency $\omega$, and define
	\begin{align}
		\tilde{s}=e^{i\omega t}s
	\end{align}
	and $\tilde{s}^x=\tilde{s}+\tilde{s}^*$, $\tilde{s}^y=i(\tilde{s}-\tilde{s}^*)$, as well as $\tilde{J}^\mu=\int d{\bf x}\int d{\bf p}\eta({\bf x})\tilde{s}^\mu$ for $\mu\in\{x,y\}$, accordingly. This frame is chosen such that the steady state $\tilde s_0 =e^{i\omega t}s_0$ is time-independent, i.e.,
	\begin{align}
		\frac{d\tilde{s}_0}{dt}=\left[i\omega s_0 + \frac{d{s}_0}{dt}\right]e^{i\omega t}=0
		.
	\end{align}We now consider small fluctuations $\delta \tilde{\bf s}=(\delta \tilde{s}^x,\delta \tilde{s}^y,\delta \tilde{s}^z)^T$ around the stationary solutions that we can parameterize by
	\begin{align}
		\tilde{s}^x_0=&\rho\cos(\psi)\sin(K),\nonumber\\
		\tilde{s}^y_0=&\rho\sin(\psi)\sin(K),\nonumber\\
		\tilde{s}^z_0=&\rho\cos(K).
		\label{Sphericalcoordinatessolution}
	\end{align}
	In this rotating frame, we also keep the convention introduced in Eq.~\eqref{Jpar0} and Eq.~\eqref{zero}
	\begin{align}
		\label{Jpar02}
		J_0^{\parallel}=&\int d{\bf x}\int d{\bf p}\,\eta \tilde{s}^x_0,\\
		0=&\int d{\bf x}\int d{\bf p}\,\eta\tilde{s}^y_0,
	\end{align}
	meaning that the collective dipole is chosen to be always pointing in the $x$ direction. By linearizing and solving the equations for $\delta \tilde{\bf s}$ we find linear equations for $\delta\tilde{\bf J}=(\delta \tilde{J}^x,\delta \tilde{J}^y)^T$. We find then that the time evolution is described by
	\begin{align*}
		\delta\tilde{\bf J}\propto e^{\mathrm{\nu}_0t},
	\end{align*}
	where $\nu_0$ is the zero of the dispersion relation
	\begin{align}
		\label{DispersionSR}
		D_{\mathrm{SR}}(\nu)=\mathrm{det}\left[{\bf D}(\nu)\right]
	\end{align}
	with the largest real component. We present a detailed derivation and the actual form of the dispersion relation in Appendix~\ref{App:DispSR}. This dispersion relation can be used to determine the nature of the instability of the superradiant configuration. Specifically, for a particular example that we study later in Sec.~\ref{sec:6}, we will show that the amplification of fluctuations occurring for $\mathrm{Re}(\nu_0)>0$ can lead to a transition to a multicomponent superradiant emission regime. 
	
	After providing all the theory that is required to analyze the beam-cavity system, we will analyze in the next section a specific model where we apply all the results of Sec.~\ref{sec:3} and~\ref{sec:4}.
	
	\section{\label{sec:5}An atomic beam with a single velocity traversing an off-resonant optical cavity} 
	We will now investigate a system consisting of an atomic beam composed of atoms with an identical velocity ${\bf v} = (v_x, 0, 0)^T$ travelling across one antinode of the cavity mode (see Fig.~\ref{Fig:1}). We assume that the cavity mode can be modeled by
	\begin{align}
		\label{eta}
		\eta({\bf x})=\Theta(x+w)-\Theta(x-w),
	\end{align} 
	which simplifies the cavity profile to a box with length $2w$, where $w$ is the waist of the cavity mode. The transit time $\tau$ is thus fixed to be $\tau=2w/v_x$. For $t>\tau$, the corresponding homogeneous density of atoms is given by
	\begin{align}
		\label{rho}
		\rho=\frac{N}{2w}.
	\end{align}
	
	\subsection{Non-superradiant phase}
\begin{figure}[h!]
		\center
		\includegraphics[width=1\linewidth]{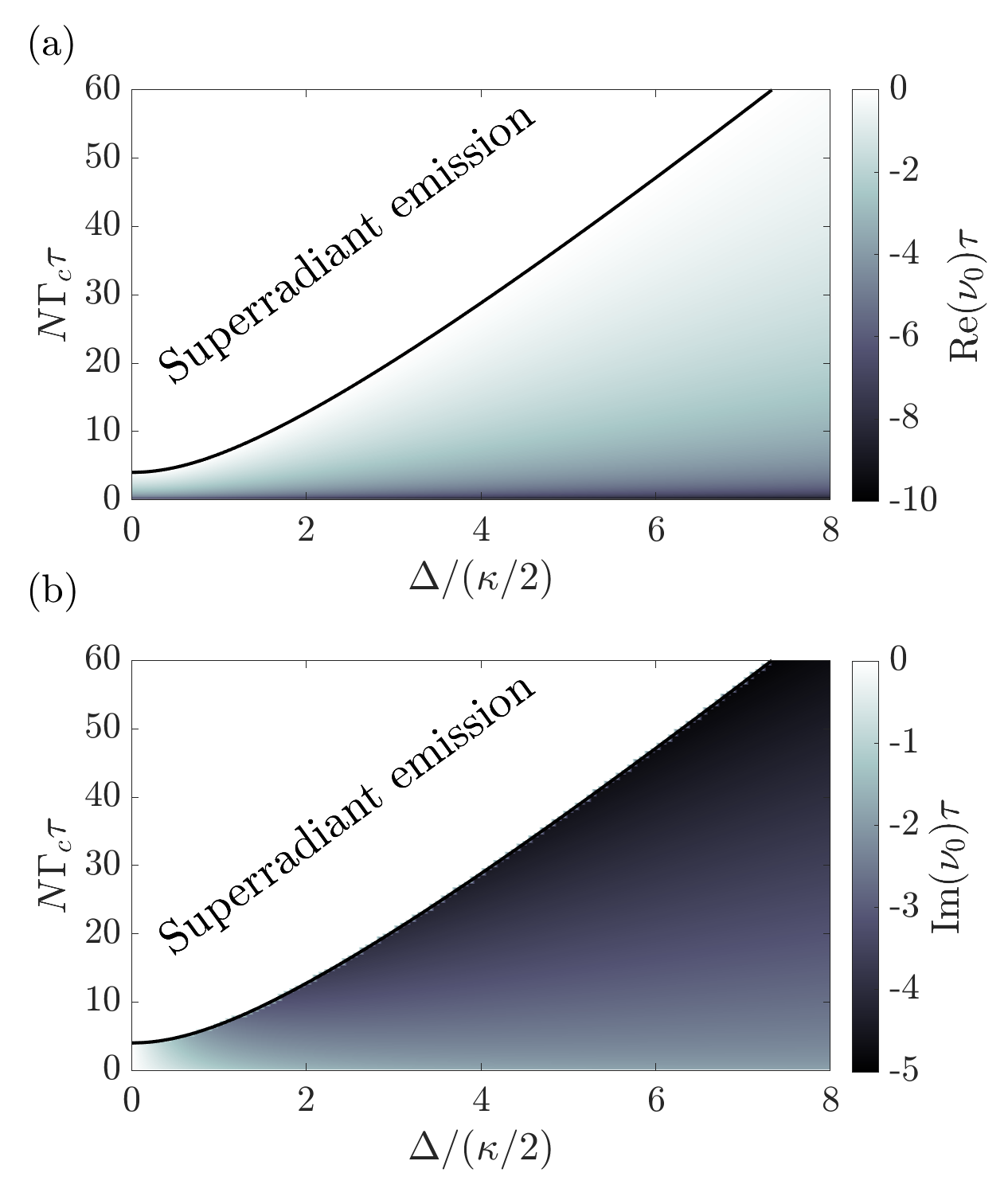}
		\caption{(a) The real component $\mathrm{Re}(\nu_0)$ and (b) the imaginary component $\mathrm{Im}(\nu_0)$ of the zero $\nu_0$ with the largest real component of $D(\nu)$ in Eq.~\eqref{Dexact}. They are plotted as a function of the detuning $\Delta$ in units of $\kappa/2$ and the collective linewidth $N\Gamma_c$ in units of $1/\tau$. The black solid line is determined by $\mathrm{Re}(\nu_0)=0$ above which we expect superradiant emission.\label{Fig:3}}
	\end{figure}

	We will first determine the stability of the non-superradiant configuration given by Eq.~\eqref{nonsupstate}. Using Eq.~\eqref{eta} and Eq.~\eqref{rho}, we can explicitly calculate the dispersion relation $D(\nu)$ given in Eq.~\eqref{Dispersionrelationnonsupp} that takes the form
	\begin{align}
		\label{Dexact}
		D(\nu)=&1-\frac{N\Gamma_c\tau}{2}\cos(\chi)
		e^{-i\chi}\frac1{\nu\tau}\left(1-\frac{1-e^{-\nu\tau}}{\nu\tau}\right).
	\end{align}
	We then numerically find the solution $\nu_0$ of $D(\nu_0)=0$ with the largest real component. In Fig.~\ref{Fig:3}, we show the real component $\mathrm{Re}(\nu_0)$ in subplot (a) and the imaginary component $\mathrm{Re}(\nu_0)$ in subplot (b) as a function of $N\Gamma_c\tau$ and of $\Delta/(\kappa/2)$, respectively.
	In Fig.~\ref{Fig:3}(a), we observe $\mathrm{Re}(\nu_0)<0$ for sufficiently small $N\Gamma_c\tau$ or large enough $\Delta/(\kappa/2)$. The solid black line marks the phase transition threshold below which the non-superradiant configuration is stable, and above which we expect superradiant emission. Specifically, for $\Delta/(\kappa/2)=0$ this threshold is given by $N\Gamma_c\tau=4$, which means that superradiant emission is only possible if the collective linewidth $N\Gamma_c$ is essentially larger than the transit time broadening $1/\tau$. In Fig.~\ref{Fig:3}(b), we plot the imaginary component $\mathrm{Im}(\nu_0)$ which is the frequency of the atomic emission relative to the atomic resonance frequency $\omega_a$. Therefore it is clear that $\mathrm{Im}(\nu_0)=0$ for $\Delta=0$,  implying that the atomic frequency, the cavity frequency, and the emission frequency are all equal. When $\Delta\neq0$, the emission frequency depends not only on $\Delta/(\kappa/2)$ but also on $N\Gamma_c\tau$.
	
	\subsection{Superradiant phase} 
	We will now study the superradiant configuration as shown in Fig.~\ref{Fig:3} above the phase transition threshold. For this we need to solve Eq.~\eqref{psi} and Eq.~\eqref{K} given Eq.~\eqref{eta} and Eq.~\eqref{rho}. Using the substitution
	\begin{align}
		\sin(\psi-\chi)=\frac{\Psi}{\sin(K)}\label{psiPsi}
	\end{align}
	we derive the differential equation
	\begin{align}
		-\omega\sin(K)v_x\frac{\partial K}{\partial x}=\frac{\Gamma_c\cos(\chi) J_0^{\parallel}}{2}\eta v_x\frac{\partial \Psi}{\partial x}.
	\end{align}
	This equation implies
	\begin{align}
		\label{Psi}
		\Psi = [1-\cos(K)]f
	\end{align}
	with
	\begin{align}
		f=-\frac{2\omega}{\Gamma_c \cos(\chi) J^{\parallel}_0},
	\end{align}
	where we have used the fact that $\eta$ is unity for $-w\leq x\leq w$ by Eq.~\eqref{eta}. Combining this result with Eq.~\eqref{psiPsi} and then solving Eq.~\eqref{K} we obtain
	\begin{align}
		\label{Kdyn}
		\sin\left[\frac{K(x)}{2}\right]=\frac{\sin\left[\frac{\sqrt{1+f^2}\Gamma_c\cos(\chi) J_0^{\parallel}(x+w)}{4v_x}\right]}{\sqrt{1+f^2}}.
	\end{align}
	
	We have now found the solutions for $K$ and $\psi$ and will use them to determine the frequency $\omega$ and the collective dipole $J_0^{\parallel}$. Using the results for $\psi$ and $K$ in Eq.~\eqref{Jpar0} and Eq.~\eqref{zero}, after some algebra we find
	\begin{align}
		\label{xicalc}
		\xi=&N\Gamma_c\tau\frac{\sin^2\left(\frac{\xi}{2}\right)}{\xi},\\
		\label{fcalc}
		-\xi\tan(\chi)=&\frac{N\Gamma_c\tau}{2}\frac{f}{\sqrt{1+f^2}}\left[1-\frac{\sin(\xi)}{\xi}\right],
	\end{align}
	with
	\begin{align}
		\xi=\frac{\sqrt{1+f^2}\Gamma_c\cos(\chi) J_0^{\parallel}\tau}{2}.
	\end{align}
	Given a value of $\Delta/(\kappa/2)$ and $N\Gamma_c\tau$, we can now numerically determine $\xi$ and $f$ and then calculate $J_0^{\parallel}$ and $\omega$. These values can then be used to derive $K(x)$ and $\psi(x)$. 
	
\begin{figure*}[t]
		\center
		\includegraphics[width=0.95\linewidth]{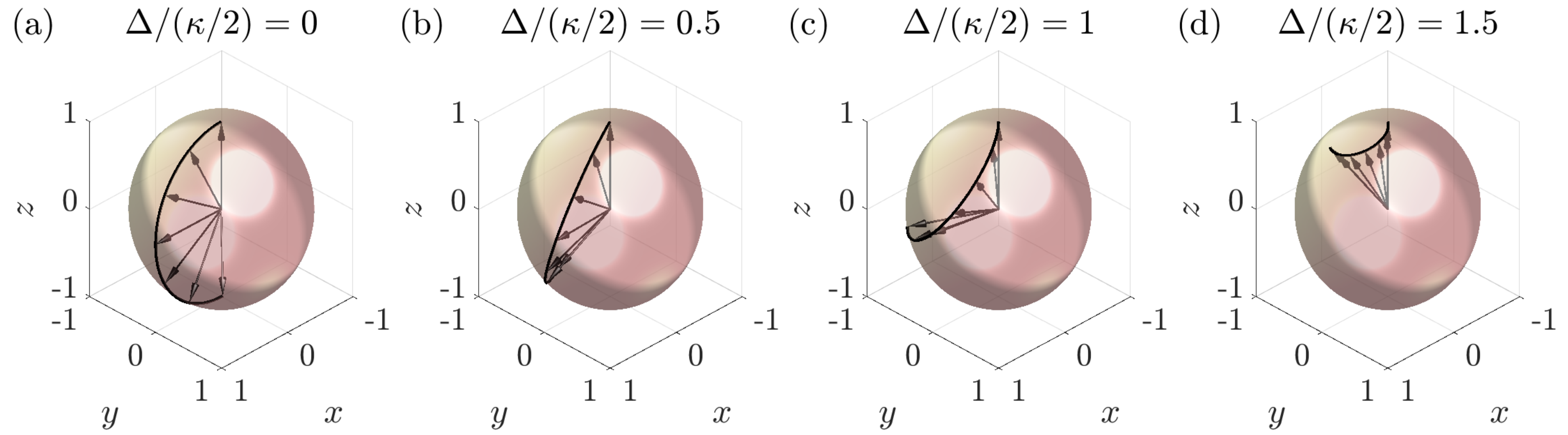}
		\caption{Bloch vectors parametrized according to Eq.~\eqref{Sphericalcoordinatessolution} where we have combined Eq.~\eqref{xicalc} and Eq.~\eqref{fcalc} to calculate $\omega$ and $J_0^{\parallel}$ and then Eq.~\eqref{psiPsi}, Eq.~\eqref{Psi}, and Eq.~\eqref{Kdyn} to calculate $K(x)$ and $\psi(x)$. The black solid lines are the traces of the Bloch vectors for $-w\leq x\leq w$. We have used $N\Gamma_c\tau=10$ and four different values of $\Delta/(\kappa/2)$ [see titles of subplots (a)--(d)] for the numerical values used. \label{Fig:4}}
	\end{figure*}

	In Fig.~\ref{Fig:4} we show the result for four different values of $\Delta/(\kappa/2)$ with a fixed $N\Gamma_c\tau=10$ where we derive $K(x)$ and $\psi(x)$ and then use Eq.~\eqref{Sphericalcoordinatessolution} to illustrate the dynamics of the dipoles on the Bloch sphere (see Fig.~\ref{Fig:2} with $\phi=\psi$) for $-w\leq x\leq w$. We have normalized the Bloch vector to length unity.
	Since the atoms enter in the excited state $|e\rangle$, the Bloch vector is pointing along the $z$ direction initially for $x=-w$. For all cases the collective dipole $J_0^{\parallel}$, which is here determined by the integral of all the Bloch vectors along the trajectory for $-w\leq x\leq w$, points in the $x$ direction by choice [see Eq.~\eqref{zero} and Eq.~\eqref{Jpar02}]. In Fig.~\ref{Fig:4}(a) where $\Delta=0$, the Bloch vector remains in a plane that is spanned by the $z$ axis and the collective dipole. This is different for non-vanishing $\Delta$ values [see Fig.~\ref{Fig:4}(b--d)] where the Bloch vectors leave this plane. We observe that the total curve becomes shorter for increasing $\Delta$ values and the length of the collective dipole also decreases for these parameters.
	
\begin{figure}[h!]
		\center
		\includegraphics[width=1\linewidth]{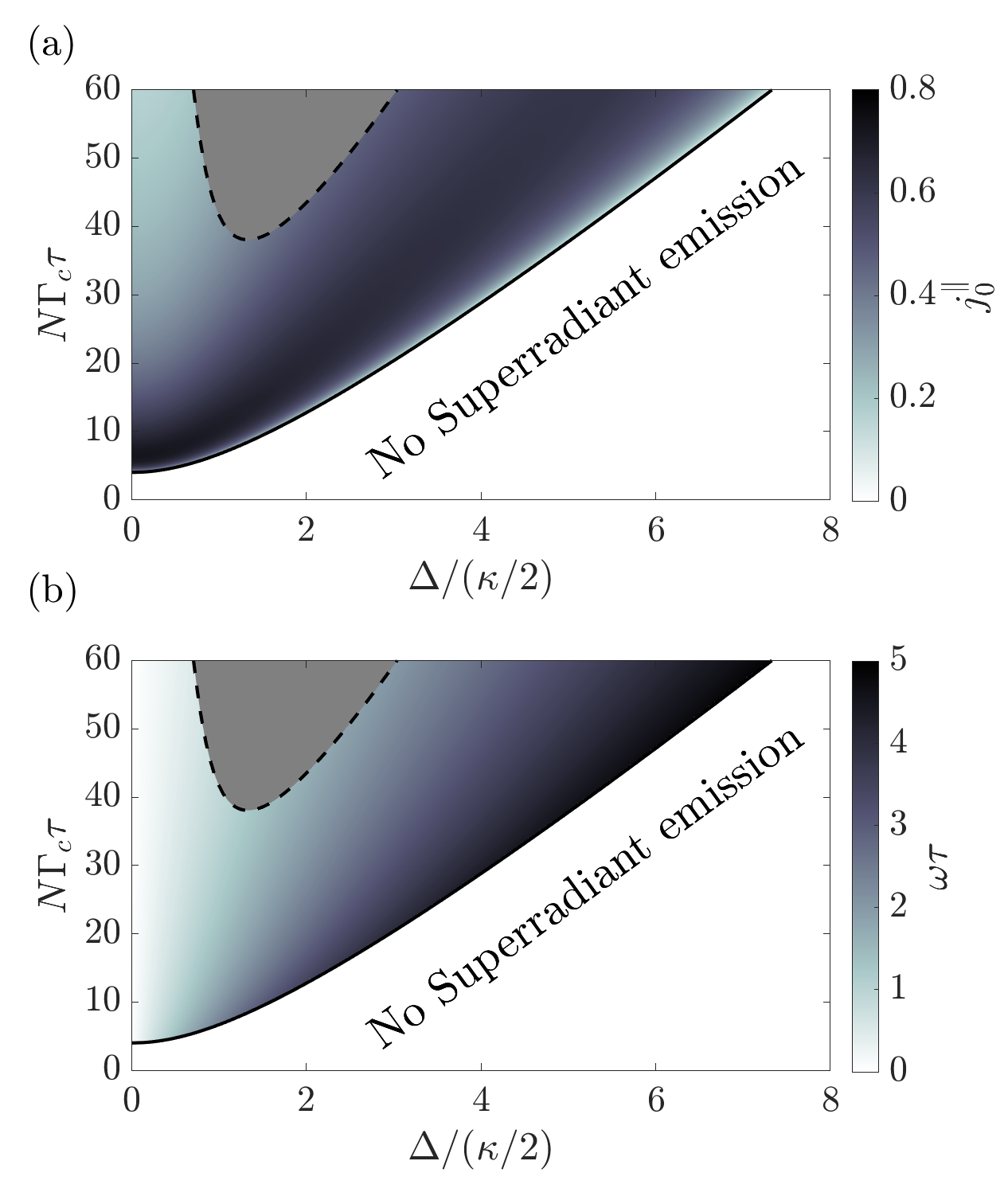}
		\caption{The normalized collective dipole $j_0^{\parallel}=J_0^{\parallel}/N$ (a) and the frequency $\omega$ in units of $1/\tau$ (a) as a function of $\Delta/(\kappa/2)$ and the collective linewidth $N\Gamma_c$ in units of $1/\tau$. The results are calculated using Eq.~\eqref{xicalc} and Eq.~\eqref{fcalc}. The black dashed line is the boundary of the gray area where the superradiant configuration transitions to a multicomponent superradiant regime. This has been determined using the solution of $j_0^{\parallel}$ and $\omega$ to find zeros of the dispersion relation in Eq.~\eqref{DispersionSR}.\label{Fig:5}}
	\end{figure}

	In order to study this effect, in Fig.~\ref{Fig:5}(a) we show the normalized collective dipole $j_0^{\parallel}=J_0^{\parallel}/N$ for different values of $N\Gamma_c\tau$ and $\Delta/(\kappa/2)$. We observe the same transition threshold between the superradiant and non-superradiant phases as shown by the black solid line in Fig.~\ref{Fig:3}. This transition is continuous but not differentiable.   
	Above the threshold, we find a non-vanishing value for the collective dipole. In Fig.~\ref{Fig:5}(b) we show the value of the frequency $\omega$ that has been calculated for the same parameter regime as $j_0^{\parallel}$ in Fig.~\ref{Fig:5}(a). We see that $\omega$ vanishes for $\Delta=0$ which implies that the atomic frequency $\omega_a$, the cavity frequency $\omega_c$, and $\omega$ are equal. For a given value of $N\Gamma_c\tau$ the frequency $\omega$ increases linearly with $\Delta/(\kappa/2)$. This shows that the cavity pulling coefficient $\mathcal{P}=\omega/\Delta$ in the superradiant regime is independent of $\Delta$ even for large values of $\Delta/(\kappa/2)$.
	
	We have also derived the stability of the superradiant configuration using the dispersion relation in Eq.~\eqref{DispersionSR}. We have found zeros $\nu_0$ with positive real part for the parameter region that is shown as gray area in Fig.~\ref{Fig:5} bounded by a black dashed line. This is the parameter space where we expect a different dynamical phase because the stationary superradiant and the non-superradiant solutions are unstable. 
	
	We will now compare our analytical finding with numerical simulations.
	
	\subsection{Numerical study}
	We numerically integrate Eq.~\eqref{sxdensityel}--\eqref{szdensityel} using the mode function in Eq.~\eqref{eta} and the homogeneous density in Eq.~\eqref{rho}.
	
	\subsubsection{Superradiant to non-superradiant regime}
\begin{figure}[h!]
		\center
		\includegraphics[width=1\linewidth]{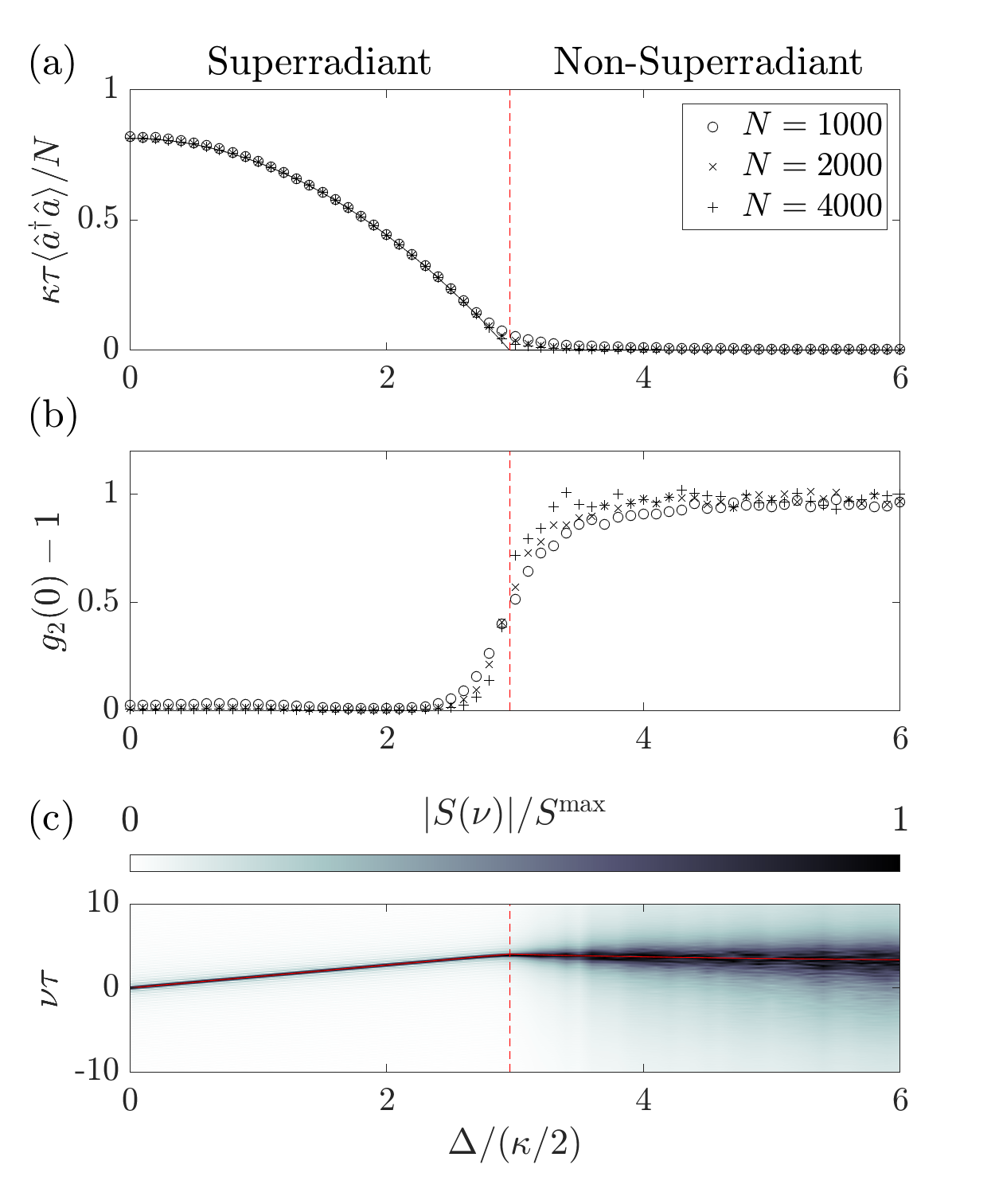}
		\caption{(a) The cavity output power $\kappa\langle \hat{a}^{\dag}\hat{a}\rangle$ in units of $N/\tau$ [see Eq.~\eqref{outputpower}] and (b) the value of $g_2(0)-1$ [see Eq.~\eqref{g2}] as functions of $\Delta/(\kappa/2)$ for various values of $N$ [see inset of subplot (a)].  (c) The spectrum $|S(\nu)|$ [see Eq.~\eqref{spectrum}] normalized for every value of $\Delta/(\kappa/2)$ by the maximum $S^\mathrm{max}=\max_\nu|S(\nu)|$  as a function of $\nu$ in units of $1/\tau$ and of $\Delta/(\kappa/2)$ obtained by numerically integrating Eqs.~\eqref{sxdensityel}--\eqref{szdensityel} for $N=4000$. For all simulations we have used $N\Gamma_c\tau=20$. The black solid line in subplot (a) is calculated from the solution $j_0^{\parallel}$ obtained from Eq.~\eqref{xicalc} and Eq.~\eqref{fcalc}. The vertical red dashed lines mark the analytical threshold between the superradiant and non-superradiant emission regimes. For (c) we have used $t_0=10\tau$ and $t_{\mathrm{cut}}=20\tau$. The red solid line in (c) in the superradiant regime is the frequency $\omega$ calculated using Eq.~\eqref{xicalc} and Eq.~\eqref{fcalc}. The red solid line in (c) in the non-superradiant regime is $\mathrm{Im}(\nu_0)$ where $\nu_0$ is the zero of Eq.~\eqref{Dispersionrelationnonsupp} with the largest real part. All simulations have been performed for a total time $T=200\tau$ and averaged over $100000/N$ different initializations.\label{Fig:6}}
	\end{figure}

	We first investigate the crossover regime from the superradiant to the non-superradiant phase for a fixed $N\Gamma_c\tau=20$ and various values of $\Delta/(\kappa/2)$ and $N$. Figure~\ref{Fig:6}(a) shows the cavity output power in units of $N/\tau$. This quantity can be interpreted as the number of photons that are emitted per atom during the transit time~$\tau$. It is calculated from
	\begin{align}
		\label{outputpower}
		\frac{\kappa\langle \hat{a}^{\dag}\hat{a}\rangle}{N/\tau}=&\Gamma_c\cos^2(\chi)\tau\frac{\langle J^*J\rangle}{N}, 
	\end{align}
	where we have used Eq.~\eqref{eliminatedalpha} and $J=(J^x-iJ^y)/2$ is taken from the numerical integration. On the other hand we can take our analytical results where we expect $\langle J^*J\rangle=N^2j_0^{\parallel}/4$ to predict the cavity output power. In Fig.~\ref{Fig:6}(a), we show the numerical results of the output power as dotted lines with different markers which indicate different atom numbers (see inset). The analytical results calculated from $j_0^{\parallel}$ is shown as the solid black line. We find very good agreement of the numerical and analytical results for all parameters. In general we observe that at the transition from the superradiant to the non-superradiant phase (dashed vertical red line), finite size effects smooth out the non-analyticity, which is expected from the analytical results. 
	
	To study the coherence properties we also investigate the second-order Glauber $g_2$ function defined as
	\begin{align}
		\label{g2}
		g_2(0)=\frac{\langle J^{*}JJ^{*}J\rangle}{\langle J^{*}J\rangle^2},
	\end{align}
	which is shown in Fig.~\ref{Fig:6}(b). Well inside the superradiant phase we observe $g_2(0)\approx 1$, which indicates second-order coherent light. This result is as expected because in this regime and for large intracavity atom number $N$, the collective dipole is coherent and therefore noise only plays a minor role. As a consequence we can use $\langle J^{*}J\rangle\approx (Nj_0^{\parallel}/2)^2$ and $\langle J^{*}JJ^{*}J\rangle \approx \langle J^{*}J\rangle^2$. The value of $g_2(0)$ increases at the threshold and reaches $g_2(0)\approx 2$ well inside the non-superradiant regime. This result indicates thermal light. 
	
	In order to have access to the emission frequency of the cavity field we have also calculated the spectrum
	\begin{align}
		\label{spectrum}
		S(\nu)=\int_{0}^{t_{\mathrm{cut}}}dt e^{-i\nu t}\langle J^*(t+t_0)J(t_0)\rangle,
	\end{align}
	where $t_0\gg\tau$ is a time after which we expect the system to reach stationary state and $t_{\mathrm{cut}}$ is a numerical integration time. This spectrum is shown in Fig.~\ref{Fig:6}(c) as a function of the frequency $\nu$ in units of $1/\tau$ and for different values of $\Delta/(\kappa/2)$. We have normalized this spectrum for every value of $\Delta/(\kappa/2)$ such that $|S(\nu)|\leq1$. In the superradiant phase we observe a narrow peak of the spectrum. Specifically, the peak is centered at $\omega=0$ for $\Delta=0$. For increasing values of $\Delta/(\kappa/2)$ from zero, we find a linear increase of the emission frequency described by this peak. The red solid line in the superradiant regime indicates the analytical solution of $\omega$ that has been presented in Fig.~\ref{Fig:5}(b) and is in very good agreement with the numerical results. The linear behavior of the emission frequency is determined by the pulling coefficient, $\omega=\mathcal{P}\Delta$, where we find $\mathcal{P}\kappa\tau\approx2.8$.
	
	In the non-superradiant regime we observe a much broader spectrum and also a different behavior of the emission frequency. The red solid line in
	the non-superradiant regime describes the solution $\mathrm{Im}(\nu_0)$ shown in Fig.~\ref{Fig:3}(b). We find good agreement between this solution and the peak of the spectrum in the non-superradiant phase.
	
	\subsubsection{Stationary to multicomponent superradiant regime}
\begin{figure}[h!]
		\center
		\includegraphics[width=1\linewidth]{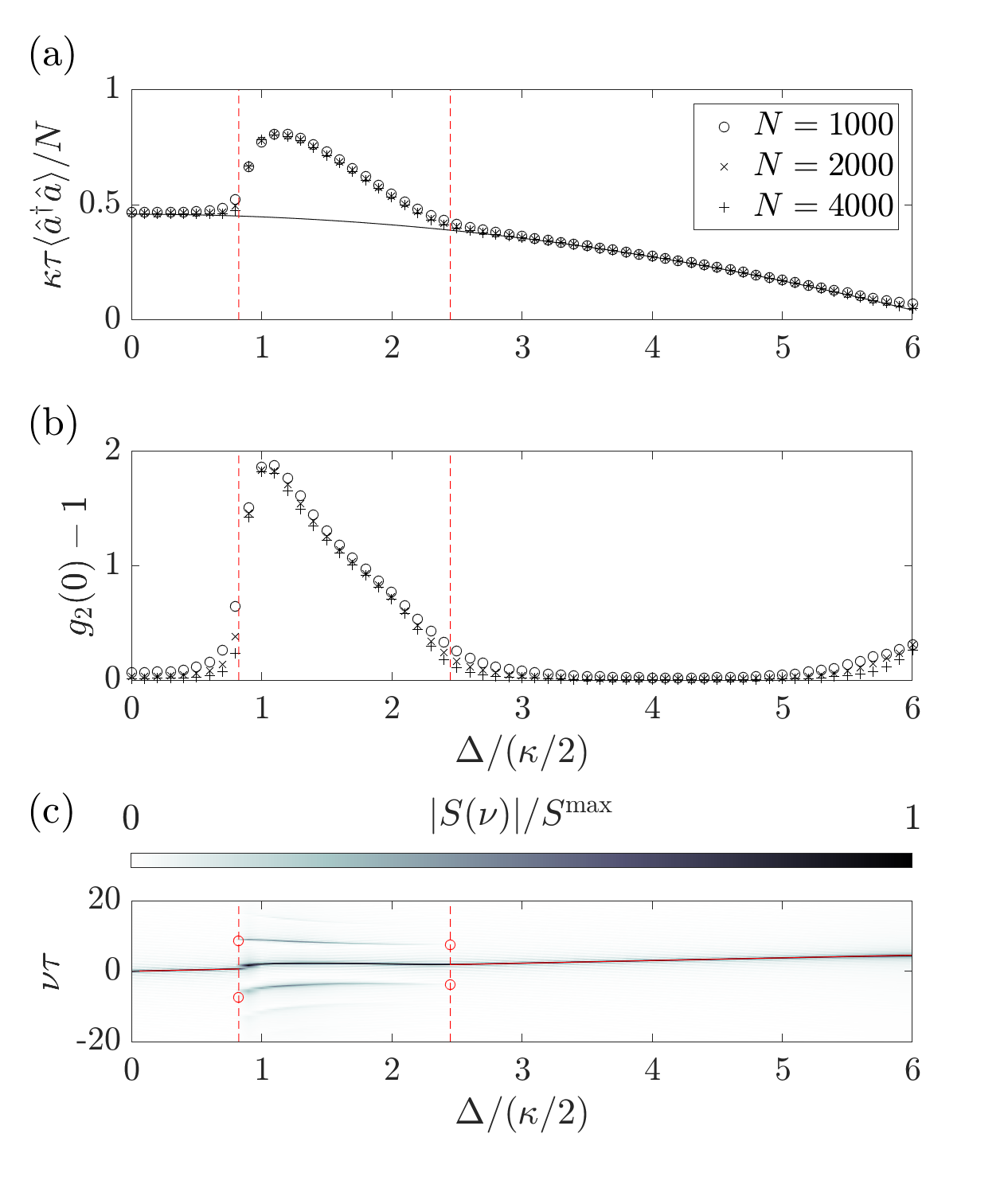}
		\caption{(a) The cavity output power $\kappa \langle \hat{a}^{\dag}\hat{a}\rangle$ in units of $N/\tau$ [see Eq.~\eqref{outputpower}] and (b) the value of $g_2(0)-1$ [see Eq.~\eqref{g2}] as functions of $\Delta/(\kappa/2)$ for various values of $N$ [see inset of subplot (a)].  (c) The spectrum $|S(\nu)|$ [see Eq.~\eqref{spectrum}] normalized for every value of $\Delta/(\kappa/2)$ by the maximum $S^\mathrm{max}=\max_\nu|S(\nu)|$ as a function of $\nu$ in units of $1/\tau$ and of $\Delta/(\kappa/2)$ obtained by numerically integrating Eqs.~\eqref{sxdensityel}--\eqref{szdensityel} for $N=4000$. For all simulations we have used $N\Gamma_c\tau=50$. The black solid line in subplot (a) is calculated from the solution $j_0^{\parallel}$ obtained from Eq.~\eqref{xicalc} and Eq.~\eqref{fcalc}. The vertical red dashed lines border the multicomponent regime. For subplot (c) we have taken the values $t_0=10\tau$ and $t_{\mathrm{cut}}=20\tau$. The red solid line in (c) in the superradiant regime is the frequency $\omega$ calculated using Eq.~\eqref{xicalc} and Eq.~\eqref{fcalc}. The red circles in (c) at the phase thresholds are the values of $\omega\pm\mathrm{Im}(\nu_1)$, where $\nu_1$ is the zero of Eq.~\eqref{DispersionSR} with the largest real part. All simulations have been performed for a total time $T=200\tau$ and averaged over $100000/N$ different initializations.\label{Fig:7}}
	\end{figure}

	We will now investigate the transition from the stationary superradiant phase to a multicomponent superradiant phase (grey region in Fig.~\ref{Fig:5}), first along $N\Gamma_c\tau=50$ for different values of $\Delta/(\kappa/2)$. As we will show below, in this multicomponent superradiant phase we observe polychromatic superradiant emission where the spectrum shows several frequency components. We first study the output power $\kappa\tau\langle \hat{a}^{\dag}\hat{a}\rangle/N$ in Fig.~\ref{Fig:7}(a), where different markers indicate different values of $N$ (see inset). The analytical results derived from $j_0^{\parallel}$ in Eq.~\eqref{xicalc} and Eq.~\eqref{fcalc} are shown as the black solid line. The vertical dashed red lines indicate the transition from the stationary to the multicomponent superradiant region ($\Delta\lesssim1$) and from the multicomponent to the stationary region ($\Delta\lesssim2.5$). Inside the stationary superradiant phase, we find good agreement between the numerical and the analytical results. In the multicomponent regime, however, we observe that the output power spikes, indicating that every atom emits more photons than expected from the analytical theory (black solid line). 
	
	We also show the $g_2(0)$ for the same parameters in Fig.~\ref{Fig:7}(b). We find that $g_2(0)\approx1$ in the stationary superradiant regime. The slight increase for $\Delta/(\kappa/2)>5$ is due to the fact that we approach the transition to the non-superradiant regime. This can also be seen because the output power in that parameter regime approaches zero in Fig.~\ref{Fig:7}(a).
	
	In the multicomponent regime that is bordered by the two red vertical dashed lines, the $g_2(0)$ function spikes. The fact that we find values $g_2(0)>2$ indicates photon bunching in this parameter regime that cannot be explained by thermal light.
	
	The features of the emitted light are best illustrated in Fig.~\ref{Fig:7}(c) where we plot the spectrum $|S(\nu)|$ as a function of $\nu$ in units of $1/\tau$. In the stationary superradiant regime we find a narrow single peak. The position of this peak agrees very well with the frequency $\omega$ that has been calculated in Fig.~\ref{Fig:5}(b). The emission frequency follows the description $\omega=\mathcal{P}\Delta$ and we find $\mathcal{P}\kappa\tau\approx1.6$.
	
	For parameters within the region that is bordered by the two vertical red dashed lines, however, we find several narrow peaks which means that the light emission is polychromatic. The origin of the sidebands can be explained by the zero $\nu_1$ of the dispersion relation Eq.~\eqref{DispersionSR} with $\mathrm{Re}(\nu_1)>0$, signalizing an unstable superradiant configuration. The imaginary component $\mathrm{Im}(\nu_1)$ is expected to be the frequency of the sidebands relative to the central frequency $\omega$. We show $\omega\pm\mathrm{Im}(\nu_1)$ at the phase thresholds as red circles. They are in good agreement with the emerging sidebands. We emphasize that our linearized description used to calculate $\nu_1$ does not work beyond the phase thresholds to the multicomponent regime, where we need to include the full dynamical description of the atomic dipoles.
	
\begin{figure}[h!]
		\center
		\includegraphics[width=1\linewidth]{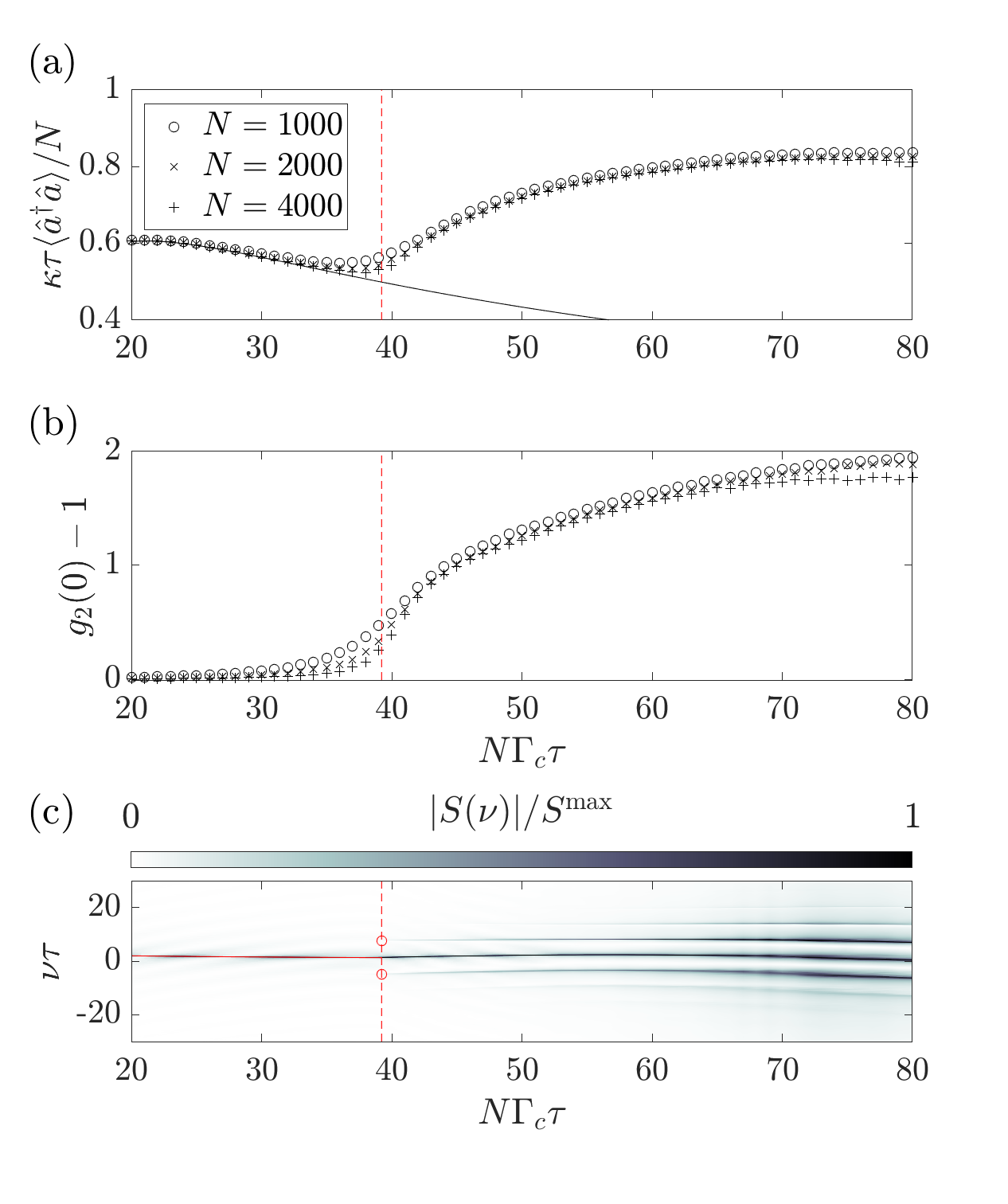}
		\caption{The same quantities as shown in Fig.~\ref{Fig:7} but for a fixed value of $\Delta/(\kappa/2)=1.5$ and as a function of $N\Gamma_c\tau$. The vertical red dashed line marks the transition from the stationary to the polychromatic superradiant regime and the red circles in (c) at the phase threshold are the values of $\omega\pm\mathrm{Im}(\nu_1)$. The remaining parameters are the same as in Fig.~\ref{Fig:7}.\label{Fig:8}}
	\end{figure}

	We have also studied the same transition for a fixed value of $\Delta/(\kappa/2)=1.5$ when we vary $N\Gamma_c\tau>20$. For these parameters, we expect the phase threshold to be around $N\Gamma_c\tau\approx40$, shown as the vertical red dashed line in Fig.~\ref{Fig:8}. In Fig.~\ref{Fig:8}(a) we show the output power for different values of $N$ using different markers (see inset). The black solid line is the analytical result calculated from $j_0^{\parallel}$. In the stationary superradiant regime the analytical and the numerical results are in good agreement. Beyond the threshold we observe an increasing value of the numerically calculated output power while the analytical result keeps decreasing.  
	
	In Fig.~\ref{Fig:8}(b) we find that the light field is second order coherent [i.e., $g_2(0)\approx1$] inside the stationary superradiant phase. When we enter the multicomponent regime we observe an increasing value of $g_2(0)$. The maximum value of $g_2(0)$ for the given parameters is close to $g_2(0)\approx3$. 
	
	The spectrum $|S(\nu)|$ is visible in Fig.~\ref{Fig:8}(c) as a function of $\nu$ in units of $1/\tau$. We find one narrow peak of the spectrum in the stationary superradiant regime. The corresponding emission frequency is in good agreement with the analytical value (red solid line) of $\omega$ calculated in Fig.~\ref{Fig:5}(b). At the transition we find two emerging sidebands. These sidebands have been compared with $\omega\pm\mathrm{Im}(\nu_1)$ (red circles), where $\nu_1$ is the zero of Eq.~\eqref{DispersionSR} with the largest real component. They are in good agreement with the numerical results. Beyond the transition point we observe an increasing number of sidebands.
	
	\subsection{Cavity pulling}
	At the end of this section we derive the cavity pulling coefficient $\mathcal{P}$ that describes the change of the emission frequency $\omega$ when the atomic transition and the cavity mode are not resonant. For this we use Eq.~\eqref{Pulling} and solve the integral in Eq.~\eqref{Cperp} using the mode function in Eq.~\eqref{eta} and the atomic density in Eq.~\eqref{rho}. Since the cavity pulling coefficient is the result for small detuning $\Delta/(\kappa/2)\ll1$, we can use Eq.~\eqref{fcalc} and find $f\propto \Delta/(\kappa/2)$ and neglect the second order in $f^2\approx0$. Consequently, we find $\xi=\Gamma_cJ_{0}^{\parallel}\tau/2$ and can use Eq.~\eqref{xicalc} to calculate $J^{\parallel}_0$. The value of $\xi$ can then be used to calculate the timescale
	\begin{align}
		C_{\perp}=\frac{N\Gamma_c\tau^2}{2}\frac{1-\frac{\sin(\xi)}{\xi}}{\xi^2}.    
	\end{align}
	
\begin{figure}[h!]
		\center
		\includegraphics[width=1\linewidth]{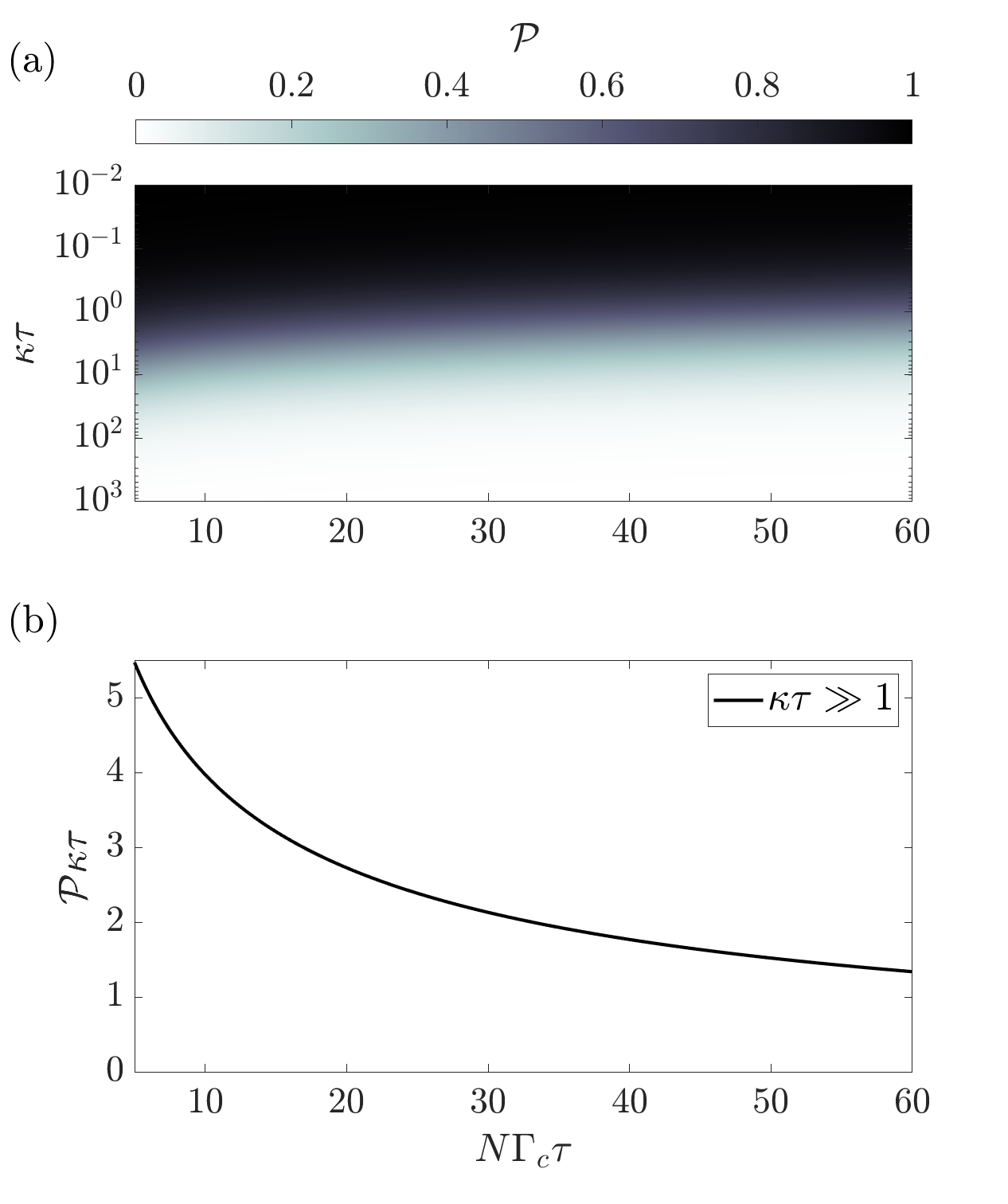}
		\caption{(a) The cavity pulling coefficient $\mathcal{P}$ defined in Eq.~\eqref{Pulling} as a function of the cavity linewidth $\kappa$ and the collective decay $N\Gamma_c$, both in units $1/\tau$. For the calculation of $\mathcal{P}$ we have solved Eq.~\eqref{Cperp} using the solution of Eq.~\eqref{xicalc} for $f=0$. (b) The cavity pulling coefficient $\mathcal{P}$ normalized by $1/(\kappa\tau)$ as a function of $N\Gamma_c$ in units $1/\tau$. For the derivation we have calculated $\mathcal{P}=\omega/\Delta$ that is independent of $\Delta$ in the limit $\kappa\tau\gg1$ where the cavity field can be eliminated. \label{Fig:9}}
	\end{figure}

	The value of $\mathcal{P}$ is shown in Fig.~\ref{Fig:9}(a) as a function of $N\Gamma_c$ and $\kappa$ both in units of $1/\tau$.
	The latter is given in a logarithmic scale to show different orders of magnitude for $\kappa\tau$. For $\kappa\tau\ll1$, the lifetime of photons is much shorter than the transit time of the atoms. In this case we expect many photons in the cavity and the resulting pulling coefficient is $\mathcal{P}\lesssim1$, showing that emission appear almost in resonance with the cavity degrees of freedom. For $\kappa\tau\gg1$, photons leave the cavity earlier than the atoms traverse the cavity. In this regime the atoms store the coherence and the frequency of the collectively emitted light is almost in resonance with the atomic transition, $\mathcal{P}\approx0$.
	
	The results obtained in the regime $\kappa\tau\gg1$ can be directly compared with our simulations. In Fig.~\ref{Fig:6}(c) and Fig.~\ref{Fig:7}(c), we have seen that the frequency $\omega$ is linear in $\Delta$ even if $\Delta/(\kappa/2)\approx1$. This is equivalent to the fact that $\mathcal{P}$ is independent on $\Delta$ in the limit $\kappa\tau\gg1$. In Fig.~\ref{Fig:9}(b), we show $\mathcal{P}$ normalized by $1/(\kappa\tau)$. This pulling coefficient is slightly different from the one that has been reported in Ref.~\cite{Liu:2020}. The reason for this discrepancy is the absence of Doppler-broadening and the cosine term in the cavity mode function in the model studied here. In fact the results in Ref.~\cite{Liu:2020} seem to be displaced by approximately a factor of $1/2$ that is due to an average over cosine-squared, and this results in a weaker effective coupling. In addition, we remark that this pulling coefficient is only valid in the stationary superradiant regime, and cannot be used for the multicomponent regime where we observe several peaks in the emission spectrum.
	
	\section{Conclusion}\label{sec:6}
	In this paper we have introduced a theoretical description for the dynamics of an atomic beam that traverses a single mode optical cavity. The atoms are described by optical dipoles with transition frequency that is detuned from the cavity frequency. We have derived the stationary phases of the atomic beam including the non-superradiant and superradiant configurations. The latter was used to calculate the cavity pulling coefficient in both the `bad' (large $\kappa$) and `good' (small $\kappa$) cavity regimes. After deriving an analytical theory for the stationary phases, we have determined the stability of the atomic dipole densities. By applying our theory to a specific model we have predicted three phases of the atomic beam. Our findings are in good agreement with numerical results where we highlight the phase transitions by examining the output power, the $g_2$ function, and also the emission spectrum. In the end we discuss cavity pulling for this specific model.
	
	The model analyzed in Sec.~\ref{sec:5} represents an idealized model since it does not capture additional relevant effects of an actual experiment such as the Doppler broadening, inhomogeneous coupling, and homogeneous broadening. However, we have shown that even such a minimal model has non-trivial solutions with monochromatic light emission and even highly dynamical phases with polychromatic light emission. Therefore we rather see this work as a stepping stone towards understanding the physics of more specific setups. Our idealized model highlights that multicomponent superradiant emission can originate from collective homogeneous frequency shifts. This work extends previous scenarios that have been studied where the dynamical phase emerges because of optomechanical effects~\cite{Jaeger:2019,Jaeger:2020} and inhomogeneous frequency shifts~\cite{Schaeffer:2020,Tang:2021,Jaeger:2021:2}. Although extensions may be necessary, the general theoretical methodology developed here will provide a good foundation for understanding any potential experimental systems. 
	
	In future, it would be interesting to understand the interplay and relation to dynamical phases that have been studied in similar atomic beam setups~\cite{Jaeger:2021:1,Jaeger:2021:2}. Moreover, while our analysis has been focusing on the light that is produced by the collective emission of the atomic beam, we have not yet investigated the atomic state in great detail. This might be especially interesting in the multicomponent superradiant regime because the dynamical character of the light field must result in a dynamical spin density. We expect that this is interesting for the study of dynamical phases and dissipative time crystals \cite{Gong:2018,Iemini:2018,Tucker:2018,Kessler:2021}.
	
	\section*{Acknowledgments}
	We acknowledge stimulating discussions with Stefan A. Sch\"affer, Athreya Shankar, and Travis L. Nicholson.
	This research is supported by the NSF AMO Grant No. 1806827; NSF PFC Grant No. 1734006; DARPA ARO Grant No. W911NF-16-1-0576; and OMA Grant No. 2016244.
	\appendix
	\section{\label{App:DispSR}Derivation of the dispersion relation for the superradiant configuration}
	In this section we will show how to calculate the dispersion relation given in Eq.~\eqref{DispersionSR}.
	
	Using Eqs.~\eqref{sxdensityel}--\eqref{sydensityel} in the frame rotating with $\omega$, the dynamics of $\delta\tilde{\bf s}$ is then governed by
	\begin{align}
		\label{fluctuations}
		\frac{\partial \delta\tilde{\bf s}}{dt}=\boldsymbol{\mathcal{L}}\delta\tilde{\bf s}+{\bf S}_0\delta{\bf \tilde{J}},
	\end{align}
	where 
	\begin{align}
		\boldsymbol{\mathcal{L}}=\mathcal{L}_0\mathbf{1}_3+\boldsymbol{\mathcal{L}}_1.
	\end{align}
	Here, we have defined
	\begin{align}
		\boldsymbol{\mathcal{L}}_1=\begin{pmatrix}
			0&\omega&\frac{\Gamma({\bf x})}{2}\cos(\chi)J^{\parallel}_0\\
			-\omega &0&\frac{\Gamma({\bf x})}{2}\sin(\chi)J^{\parallel}_0\\
			-\frac{\Gamma({\bf x})}{2}\cos(\chi)J^{\parallel}_0&-\frac{\Gamma({\bf x})}{2}\sin(\chi)J^{\parallel}_0&0
		\end{pmatrix}
	\end{align}
	and
	\begin{align}
		{\bf S}_0=\frac{\Gamma({\bf x})}{2}\begin{pmatrix}
			\cos(\chi)\tilde{s}^z_0&-\sin(\chi)\tilde{s}^z_0\\
			\sin(\chi)\tilde{s}^z_0&\cos(\chi)\tilde{s}^z_0\\
			-\cos(\chi)\tilde{s}^{x}_0-\sin(\chi)\tilde{s}^{y}_0&\sin(\chi)\tilde{s}^{x}_0-\cos(\chi)\tilde{s}^{y}_0
		\end{pmatrix}
	\end{align}
	with $\delta\tilde{\bf J}=(\delta \tilde{J}^x,\delta \tilde{J}^y)^T$. The operator $\mathcal{L}_0$ has been given in Eq.~\eqref{L0}, and $\mathbf{1}_3$ is the $3\times3$ identity matrix.
	
	The Laplace transformation of Eq.~\eqref{fluctuations} leads to
	\begin{align}
		\nu L[\delta\tilde{\bf s}]=\delta\tilde{\bf s}({\bf x},{\bf p},0)+\boldsymbol{\mathcal{L}}L[\delta \tilde{\bf s}]+{\bf S}_0L[\delta{\bf \tilde{J}}].
	\end{align}
	Now, we first solve for $L[\delta\tilde{\bf s}]$. Than we project on the first two components by multiplying with the matrix ${\mathbf{1}_{2,3}\in\mathbb{C}^{2\times 3}}$  with ones on the diagonal and zeros elsewhere. This results in two coupled equations for $L[\delta\tilde s^x]$ and $L[\delta\tilde s^y]$. 
	
	After multiplying with $\eta({\bf x})$ and integrating over the whole phase space, we arrive at
	\begin{align}
		L[\delta{\bf \tilde{J}}]=&\int d{\bf x}\int d{\bf p}\eta({\bf x})\mathbf{1}_{2,3}(\nu\mathbf{1}_3-\boldsymbol{\mathcal{L}})^{-1}\delta\tilde{\bf s}({\bf x},{\bf p},0)\nonumber\\
		&+\int d{\bf x}\int d{\bf p}\eta({\bf x})\mathbf{1}_{2,3}(\nu\mathbf{1}_3-\boldsymbol{\mathcal{L}})^{-1}{\bf S}_0L[\delta{\bf \tilde{J}}],
	\end{align}
	which can be used to solve for $L[\delta{\bf \tilde{J}}]$, resulting in
	\begin{align}
		L[\delta{\bf \tilde{J}}]={\bf D}(\nu)^{-1}\int d{\bf x}\int d{\bf p}\eta\mathbf{1}_{2,3}(\nu\mathbf{1}_3-\boldsymbol{\mathcal{L}})^{-1}\delta\tilde{\bf s}({\bf x},{\bf p},0).
	\end{align}
	where we have defined
	\begin{align}
		{\bf D}(\nu)=&{\bf 1}_2-\int d{\bf x}\int d{\bf p}\eta({\bf x})\mathbf{1}_{2,3}(\nu\mathbf{1}_3-\boldsymbol{\mathcal{L}})^{-1}{\bf S}_0\nonumber\\
		=&{\bf 1}_2-\int_{0}^{\infty}e^{-\nu t}\int d{\bf x}\int d{\bf p}\eta({\bf x})\mathbf{1}_{2,3}e^{\boldsymbol{\mathcal{L}}t}{\bf S}_0.
	\end{align}
	The dynamics of $\delta \tilde{\bf J}$ are now determined by the value of $\nu$ for which ${\bf D}(\nu)$ is not invertible.

\end{document}